# Hydrogen Development in China and the EU: A Recommended Tian Ji's Horse Racing Strategy *

Hong Xu[a,b]

[a]ETH Zurich, Switzerland, Email: hoxu@ethz.ch
[b]Technical University of Munich, Germany, Email: h.xu@tum.de

* Accepted by the 42nd *International Energy Workshop (IEW 2024)* hosted by the *International Renewable Energy Agency (IRENA)* in Bonn, Germany

**Abstract**

The global momentum towards establishing sustainable energy systems has become increasingly prominent. Hydrogen, as a remarkable carbon-free and renewable energy carrier, has been endorsed by 39 countries at COP28 in the UAE, recognizing its essential role in global energy transition and industry decarbonization. Both the European Union (EU) and China are at the forefront of this shift, developing hydrogen strategies to enhance regional energy security and racing for carbon neutrality commitments by 2050 for the EU and 2060 for China. The wide applications of hydrogen across hard-to-abate sectors and the flexibility of decentralized production and storage offer customized solutions utilizing local resources in a self-paced manner. To unveil the trajectory of hydrogen development in China and the EU, this paper proposes a comparative analysis framework employing key factors to investigate hydrogen developments in both economic powerhouses. Beyond country-wise statistics, it dives into representative hydrogen economic areas in China (Inner Mongolia, Capital Economic Circle, Yangtze River Delta) and Europe (Delta Rhine Corridor) for understanding supply and demand, industrial synergy, and policy incentives for local hydrogen industries. The derived implications offer stakeholders an evolving hydrogen landscape across the Eurasian continent and insights for future policy developments facilitating the global green transition.

*Keywords:* Hydrogen Economy, Strategy Development, Hydrogen Corridor, China, European Union

**Contents**

## 1. China, Europe, and the New Race for Hydrogen Development

Global momentum toward establishing sustainable energy systems has become increasingly prominent. In response to the United Nations (UN) Sustainable Development Goal 7 (SDG7) — "Ensuring access to affordable, reliable, and sustainable energy for all", numerous countries are anticipating to integrate renewable energy sources into their energy landscapes for mitigating environmental impacts and de-risking energy supplies. As projected by the International Energy Agency (IEA), renewables will transform the global power mix through 2027 in an accelerating pace, becoming the largest source of electricity with a share of 38%.[1]

Hydrogen, as a remarkable carbon-free and renewable energy carrier, has been endorsed by 41 governments globally in their national hydrogen strategies issued by the end of 2023.[2] Wide applications of hydrogen across hard-to-abate sectors and flexibility of de-centralised production and storage offer customised solutions for each country or region to develop their own energy strategies utilizing local resources in self-paced way. At the recent COP28 in the UAE, 39 countries jointly declared the essential role of renewable hydrogen and its derivatives in meeting global energy needs and industry decarbonization.[3]

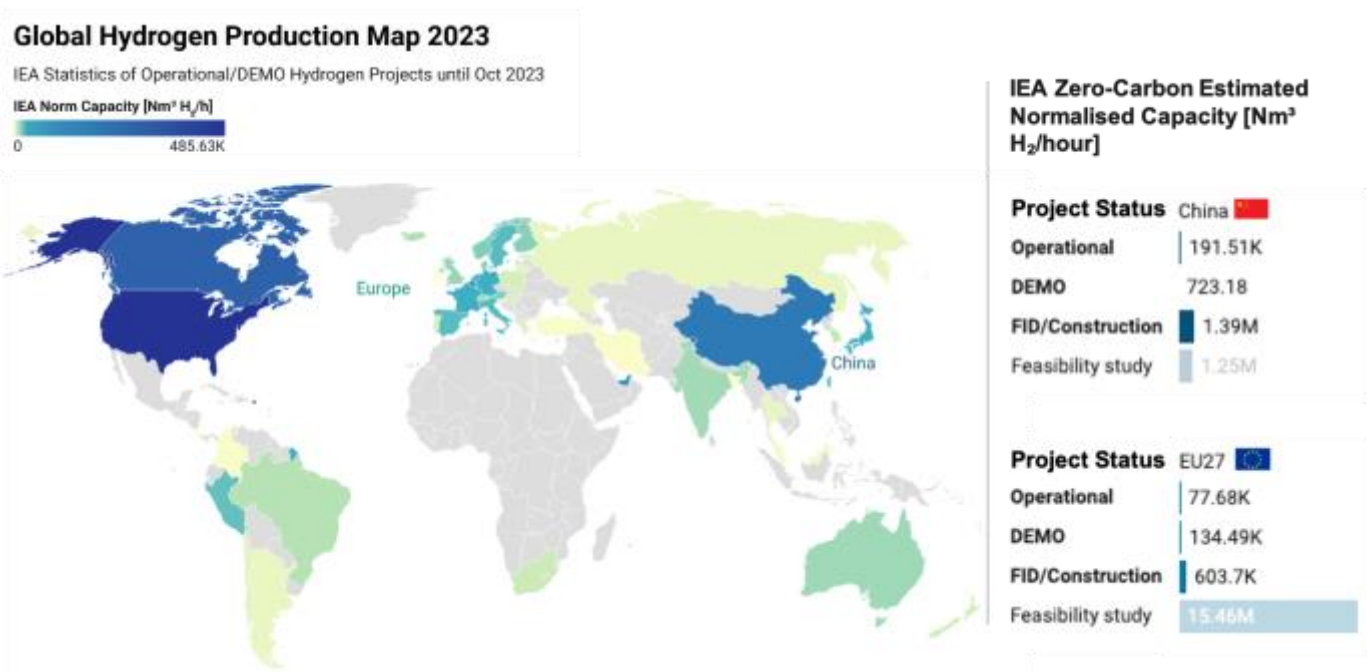

Figure 1: Global low-emission hydrogen production map (2023). Note: Final Investment Decision (FID), Demonstration Projects (DEMO).[5]

Across the Eurasian continent, China and Europe are at the forefront of this strategic shift, leading with intensively issued hydrogen policies and the rapid commissioning of hydrogen projects (Fig.1) that are shaping the trajectory of the world hydrogen economy. However, the governmental goals and industrial landscapes for hydrogen economy developments are notably differentiated by policy incentives, financial schemes, technology road-maps, industrial synergies, energy securities, and etc. There has been a surge in discussions and debates surrounding the emerging race for hydrogen development in China and Europe, mirroring the development trajectory in the photovoltaic (PV) and electric vehicle (EV) industries. Thus, it calls for in-depth comparative analysis to gain insights into the driving forces behind the hydrogen economy's momentum in Eurasia.

Over the past 20 years, series of strategic and legislative initiatives were released in China and Europe, with 2020 as a watershed year (Fig.2). China, as the world's largest hydrogen producer and consumer, rolled out hydrogen-specific national plan the first time in 2022, followed by the re-definition of hydrogen as an energy carrier, instead of hazardous chemical in its "Energy Law Act (Draft)" edited in 2020 and submitted for review in 2024.[6] The national plan aims to assert global leadership in hydrogen industry hence establish an ecosystem of diverse hydrogen applications by 2035, according to the "Medium and Long-Term Development Plan for the Hydrogen Energy Industry (2021-2035)" released by the National Development and Reform Commission (NDRC).[7] In 2024, hydrogen industry was mentioned for the first time as a frontier emerging industry in the Government Working Report released by the State Council. Large hydrogen production projects have been in construction as of 2021 in react to the guidance of national hydrogen policy, as shown by the IEA statistics in Fig.1.

The European Union (EU), as the pioneer and major investor in clean hydrogen technology, issued the comprehensive EU Hydrogen Strategy in 2020[8], along with the newly announced European Hydrogen Bank in 2023[9] and the "REPowerEU Action" Plan released in 2022. Besides, the EU recently granted a "IPCEI Hy2Intra" plan in 2024 for hydrogen infrastructures including the adaptation of 2700 km pipelines. The EU aspires to boost green hydrogen demand and production towards large-scale deployment in all sectors by 2030, ensuring secured and clean energy supply. Besides, EU planned to import similar amount of green hydrogen to fulfill the hydrogen demand by 2030. According to IEA statistics 2023, a significant share of hydrogen projects in the EU is under feasibility study compared to China, beyond large amount of on-going projects. Despite various approaches in hydrogen strategies, both China and the EU share similar environmental commitments in achieving climate neutrality by 2050 for the EU and carbon neutrality by 2060 for China.

[1] IEA (Dec 2022), Renewables 2022: Analysis and forecast to 2027. https://www.iea.org/reports/renewables-2022/executive-summary

[2] IEA (Sept 2023), Global Hydrogen Review 2023, IEA, Paris. License CC BY 4.0 https://www.iea.org/reports/global-hydrogen-review-2023

[3] COP28 (Dec 2023), COP28 Declaration of Content: Mutual Recognition of Certification Schemes for Renewable and Low-Carbon Hydrogen and Hydrogen Derivatives. https://www.cop28.com/en/cop28-uae-declaration-on-hydrogen-and-derivatives

[5] IEA (Oct 2023), Hydrogen Projects Database,https://www.iea.org/reports/hydrogen-projects-database; Data visualisation by the author with Datawrapper; Note: Final Investment Decision (FID), Demonstration Projects (DEMO).

[6] NEA (Apr 2020), "The People's Republic of China Energy Law (Draft for Comments)", https://www.nea.gov.cn/2020-04/10/c_138963212.htm (in Chinese)

[7] NDRC (Mar 2021), "The 14th Five-Year Plan and Outline of Long-term Goals for 2035", https://www.gov.cn/xinwen/2021-03/13/content_5592681.htm (in Chinese)

[8] European Commission (Jul 2020), "A hydrogen strategy for a climate-neutral Europe", https://eur-lex.europa.eu/legal-content/EN/TXT/?uri=CELEX:52020DC0301

[9] European Commission (Mar 2023), "Commission outlines European Hydrogen Bank to boost renewable hydrogen", https://energy.ec.europa.eu/news/commission-outlines-european-hydrogen-bank-boost-renewable-hydrogen-2023-03-16_en

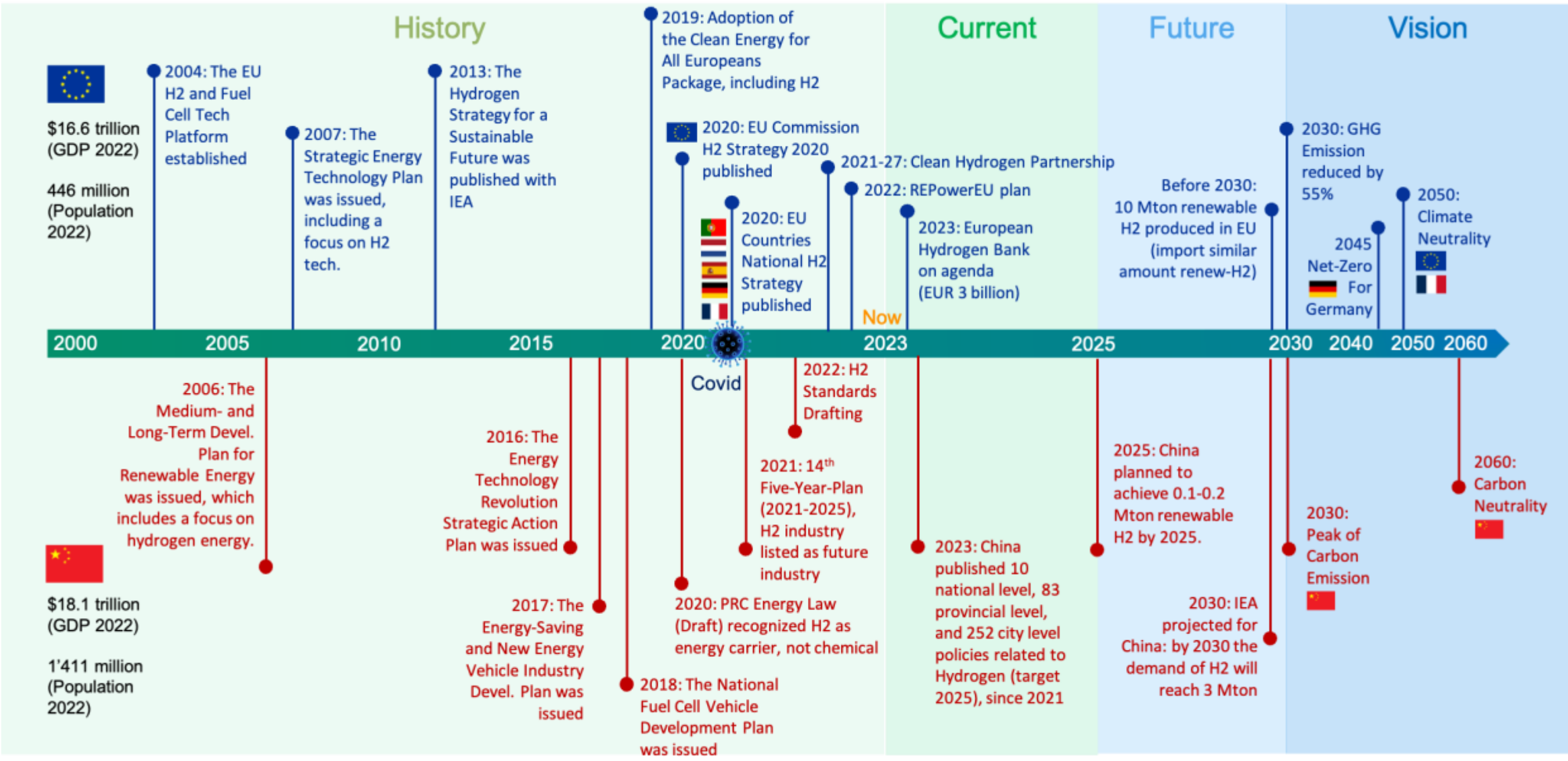

Figure 2: Timeline for hydrogen policy development in China and the EU (2000-2060)

## 2. Key Factors for Long-term Hydrogen Economy Growth

Hydrogen economy is recognized as a cornerstone for the global energy transition with rapidly growth in leading economies including China and the EU. From production to end-use sectors including electricity, transportation and etc., the hydrogen ecosystem encompasses the full spectrum of activities which can be driven by various political and economic factors. While both China and the EU announced the ambitious net-zero goals in the long-run, it is important to identify the key factors for long-term hydrogen economy growth as the measures for comparative analysis.

Hydrogen is colorless and transparent, whereas the color-labeled spectrum of hydrogen (Fig.3) are often employed for labeling the source of electricity and technology used for hydrogen production. However, it does not offer a clear notation for carbon footprint of the hydrogen, which is a key metric. The EU is now referring to "low-carbon hydrogen (LCH2)" and "Renewable hydrogen (RH2)" by setting EU taxonomy's threshold of life cycle GHG emission of 3 kg CO2(eq.)/kg H2 produced. China does not have national standard yet on hydrogen classification, whereas standard from industry group was published in 2020 and a threshold of 4.9 kg CO2(eq.)/kg H2 was listed for RH2. The LCH2, derived normally from natural gas with carbon capture, utilization, and storage (CCUS) technology, remains a transitional energy carrier. In contrast, RH2, produced through the water electrolysers using renewable sources, represents a sustainable carrier of green electricity. These two types of hydrogen can play complementary roles in the short- to long-term energy transition.

End-user sectors shape the demand of LCH2 and RH2 with different purity requirements. Presently, LCH2 is considered scalable in industries like refining, chemicals, steel, etc., meeting their high consumption needs and low-emission criteria. Looking forward, the new demand is expected to arise rapidly with the adoption of hydrogen-coupled transportation and electric grid, etc., aligning with the global expansion of RH2 production.

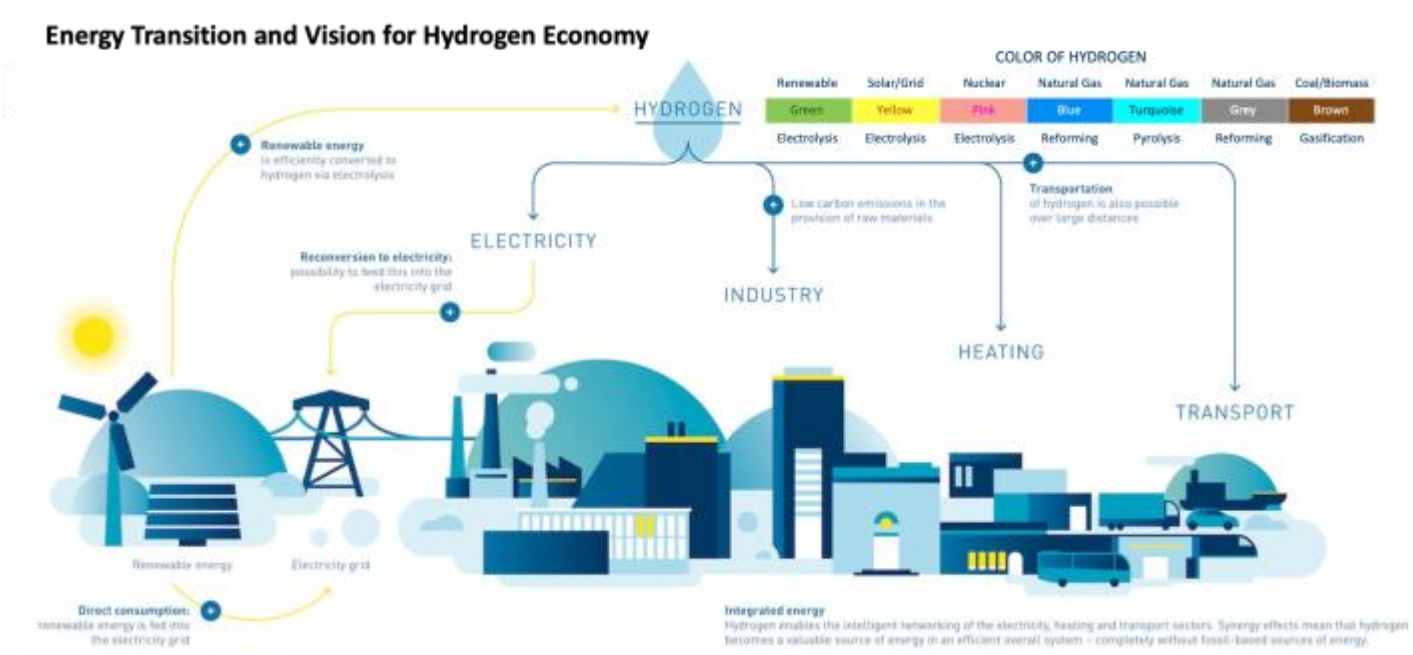

Figure 3: Energy transition and future vision for hydrogen powered economy (2020) [11]

To analyze the sustainable growth of hydrogen economy from source supply to end-users, the following key factors are considered essential. Policies and incentives are at the core of framework for long term development, which outline future market demand for RH2. The existing and planned hydrogen infrastructure, combined with technology capabilities and innovations, are fundamental to the hydrogen production and scaling up. Energy securities, along with financial supports and investments, are also shaping the distribution and economic scale of hydrogen markets.

[11] NowGmBH (2020), Vision for Hydrogen Economy, https://www.now-gmbh.de/wp\protect\discretionary{\char\hyphenchar\font}{}{}content/uploads/2020/10/Hydrogen-Economy.png

[12] Comparative analysis framework created based on research and public information collections

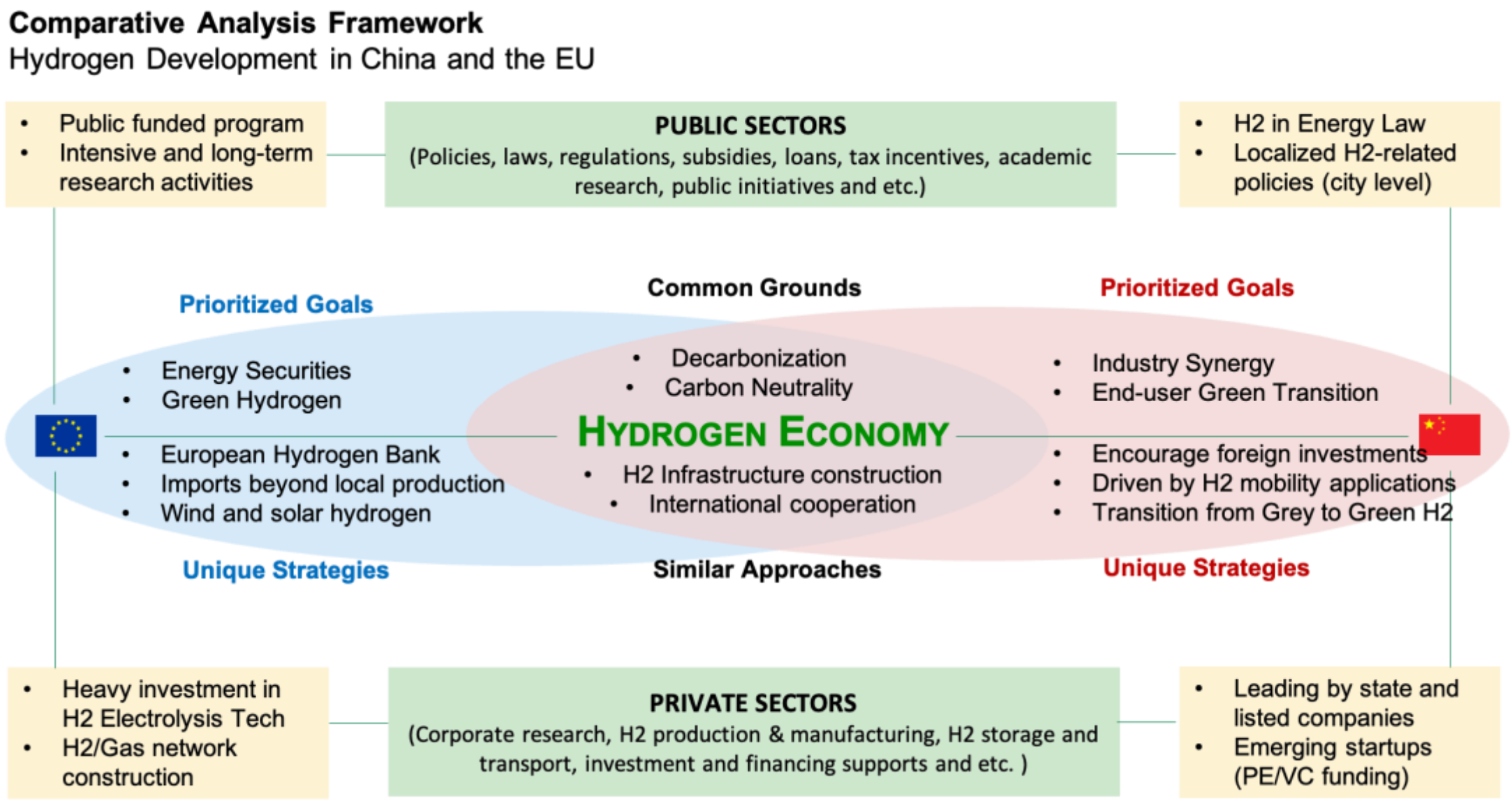

Figure 4: Comparative analysis framework for hydrogen economy in China and the EU (2023) [12]

China and the EU each bring prioritized goals and unique strategies along with common interests in decarbonisation and carbon neutrality commitments towards the development of future hydrogen economy. Given the rich renewable resources in the northern land and well-equipped industrial agglomerates in the southern cities, China prioritized the empowerment of hydrogen industry synergy and the scaling up of hydrogen production and industrial by-product utilisation in current phase. The green transition is expected when the hydrogen eco-system are well-established and is capable to supply RH2 in large production capacity.

In contrast, RH2 is at the forefront of EU's response to the challenging energy crisis since 2022. With rich mix of solar and off-shore wind resources, EU claimed the vital role of RH2 when creating market demand and scaling up production, extending this priority to the imports as well. The world first European Hydrogen Bank, is also an unique financial institution that aims to unlock immense private investments for hydrogen production. Both regions recognise the importance of hydrogen infrastructure development, such as refuelling stations and gas pipeline networks, and international cooperation.

Overarching these contexts, a comparative analysis framework (Fig.4) was developed to assess the sustainable hydrogen economy. As the core, the framework analyzes the common grounds and unique approaches in China and the EU's hydrogen strategies. The adaptions in public sectors such as laws, regulations and initiatives are considered, whereas, rapid developments in private sectors such as technologies, infrastructures, and investments are investigated.

From macro policy visions to micro hydrogen projects, this study features 3 case studies that contrasting hydrogen economic area and corridors in China and Europe, utilizing the developed framework. Although it would be challenging to cover all activities, the case studies have provided useful insights and role model for development in react to their regional hydrogen strategies. Understanding the key factors and utilizing the comprehensive framework are advantageous for stakeholders navigating the hydrogen development crossing Eurasia continent.

## 3. Common Grounds and Unique Approaches in China and EU's Hydrogen Strategies

The geography of China and Europe influences their renewable energy landscapes significantly (Fig.5). Nature-gifted Wind and solar energy are regarded as the potential major renewable sources for RH2 production. Besides, abundant water resources are necessary for electrolyser operation. The vast northern territories of China, including Inner Mongolia, Xinjiang, Beijing, and Hebei province, are natural hubs for solar and wind energy harvesting hence renewable hydrogen production. Besides, off-shore wind energy is utilized in eastern coastal cities and hydro-power is considered in river-flowed cities. In comparison, Europe's off-shore wind energy is harvested along the coastal lines of Northern France, Belgium, the Netherlands, Northern Germany, Denmark, and Norway, where they receive the strong and consistent winds from the North Sea. Navigating to the southern Europe, solar energy is the primary resource for RH2 production in Spain, Portugal, Italy and Greece.

The RH2 supply and demand are typically in line with the resource area and load centers. In China, the Heihe-Tengchong Line (Hu Line) divides the territory of China into two parts with more than 90% of populations living below the geographic line (Fig.5). This marks the general load and utility area for China's industrial and economic growth. The load centers are much diversified in Europe, given the capital and economic regions in each countries are distributed all over the continent. For major economies in Europe, Northern France, North Western Germany, Spain, Italy and Benelux Region are the population and industry dense areas.

[13]IEA, Hydrogen Projects Database (Oct 2023), https://www.iea.org/reports/hydrogen- projects- database; Data visualisation by H. Xu with Datawrapper; Note: Final Investment Decision (FID), Demonstration Projects (DEMO).

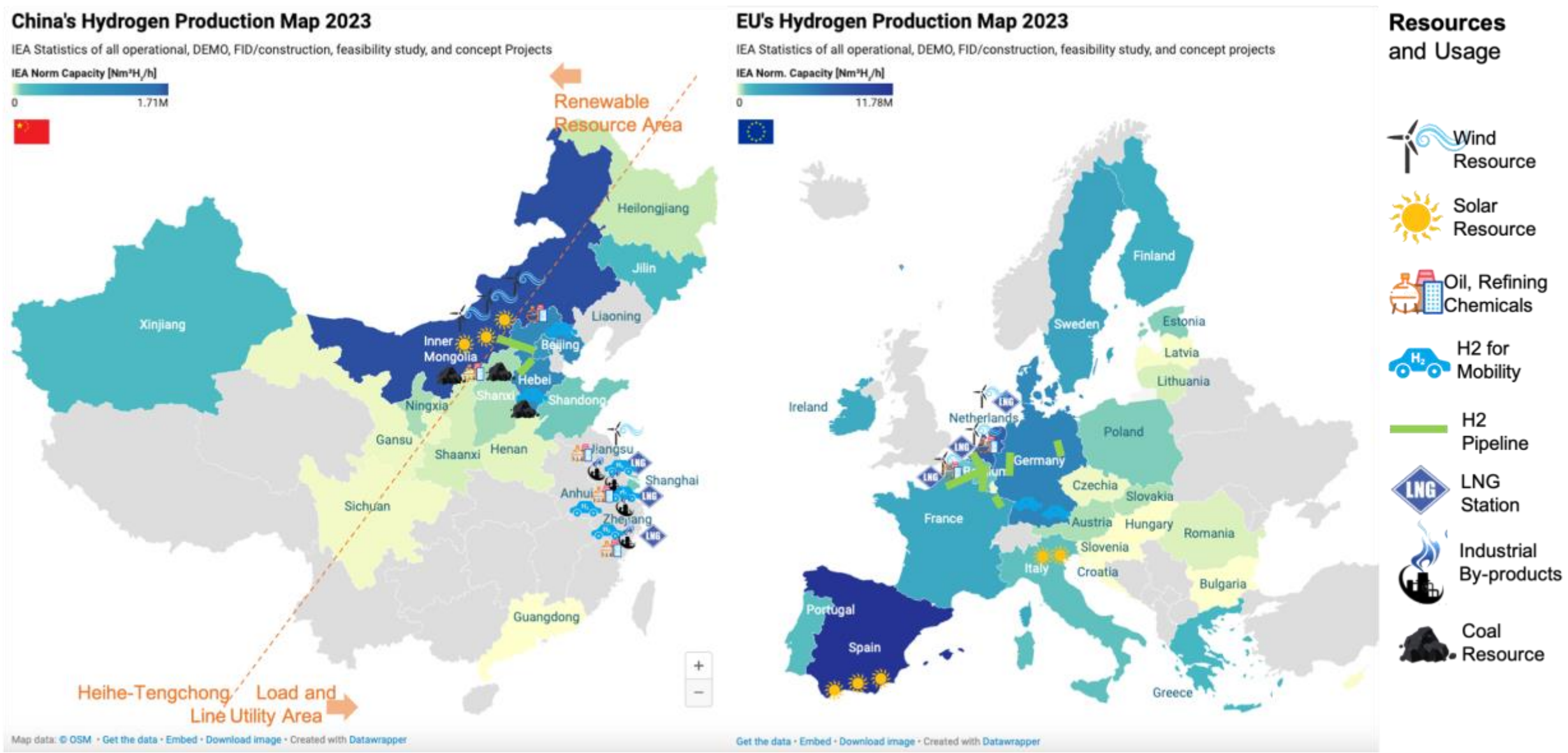

Figure 5: Hydrogen production and renewable resources in China and the EU (2023)[13]

### *3.1. China Hydrogen Strategy: Transitional Carriers and Industrial Synergies*

#### *3.1.1. China's national hydrogen strategy and adapted provincial strategies mobilising local resources and cultivating industries*

China's hydrogen industry received a comprehensive guideline upon the release of the "Medium and Long-Term Development Plan for the Hydrogen Energy Industry (2021-2035)" by the National Development and Reform Commission (NDRC) in 2022.[14] This document endorsed hydrogen energy as a critical part of China's future national energy system and an essential resource to achieve carbon neutrality goal in 2060. It also acknowledges the status of hydrogen industry as a strategically emerging industry that will drive future industrial growth in China. Towards the down-stream of hydrogen value chain, China's 13-year-long policy of subsidies for the purchase of EVs was terminated in 2022, and both the manufacturers and investors are looking for breakthrough development in FCVs, expecting a new round of mega subsidies for hydrogen mobility from the government.

The Chinese national hydrogen plan outlines three development phases (Fig.6) with specific production targets for near term and macro visions for hydrogen industry in the long-run.

In Phase 1 (2021-2025), the focus is on color-agnostic hydrogen production with specific targets for deployment of 50,000 fuel cell vehicles (FCVs) and corresponding construction of hydrogen refuelling stations. The hydrogen production goals are set to a production capacity of 0.1-0.2 Mt/y for RH2 and a reduction of 10-20 Mt/y CO2 emissions.

In Phase 2 (2025-2030), it aims to create an integrated hydrogen supply and application system. This is marked upon the successful establishment of a RH2 production and supply system, broad industrial applications of RH2, support for the carbon-peaking goal by 2030, a well-developed hydrogen technology innovation system, a reasonable layout of hydrogen production facilities, and systematic planning of the hydrogen refuelling network.

In Phase 3 (2030-2035), the goal is to establish a functional hydrogen industrial system, with an increased share of RH2 in all energy end-user sectors. Hydrogen industrial standard development, reinforced hydrogen safety supervision, and the creation of a multi-functional hydrogen energy ecosystem covering transportation, energy storage, and various industrial sectors are expected during this phase.

Since 2022, provincial regions in China have adopted the national hydrogen strategy and issued regional strategies rapidly[16], leveraging local renewable energy resources, industrial by-products for hydrogen production, and hydrogen supply chain competences. The common goal is to seeking positions of local hydrogen industry in the national industry landscape and strengthen industrial synergy among regions. 6 provinces actively published numerous hydrogen related policies as shown in Figure 6. Northern provinces such as Inner Mongolia, Hebei, and Xinjiang, with their abundance of solar and wind power resources, emphasized on the production of RH2. Meanwhile, utility and load center regions like the Beijing Capital Economic Circle (CEC) and the Yangtze River Delta (YRD) are considering a mix of LCH2 and RH2 in the initial phase, later scaling up of RH2 production. This presents the adaptability of the national strategy to provincial resources and industrial capacities. In details, the provincial hydrogen strategies are more ambitious in general than the national plan, signaling the racing of regions in China for achieving the leaderships in hydrogen industry.

[14] NDRC (Mar 2021), "The 14th Five-Year Plan and Outline of Long-term Goals for 2035" (in Chinese), https://www.gov.cn/xinwen/2021-03/13/content_5592681.htm

[15] NDRC (Mar 2021), "The 14th Five-Year Plan and Outline of Long-term Goals for 2035" (in Chinese), https://www.gov.cn/xinwen/2021-03/13/content_5592681.htm

[16] MERICS (Jun 2022), A. Brown and N. Grunberg, "CHINA'S NASCENT GREEN HYDROGEN SECTOR: How policy, research and business are forging a new industry" (in Chinese), https://www.gov.cn/xinwen/2021-03/13/content_5592681.htm

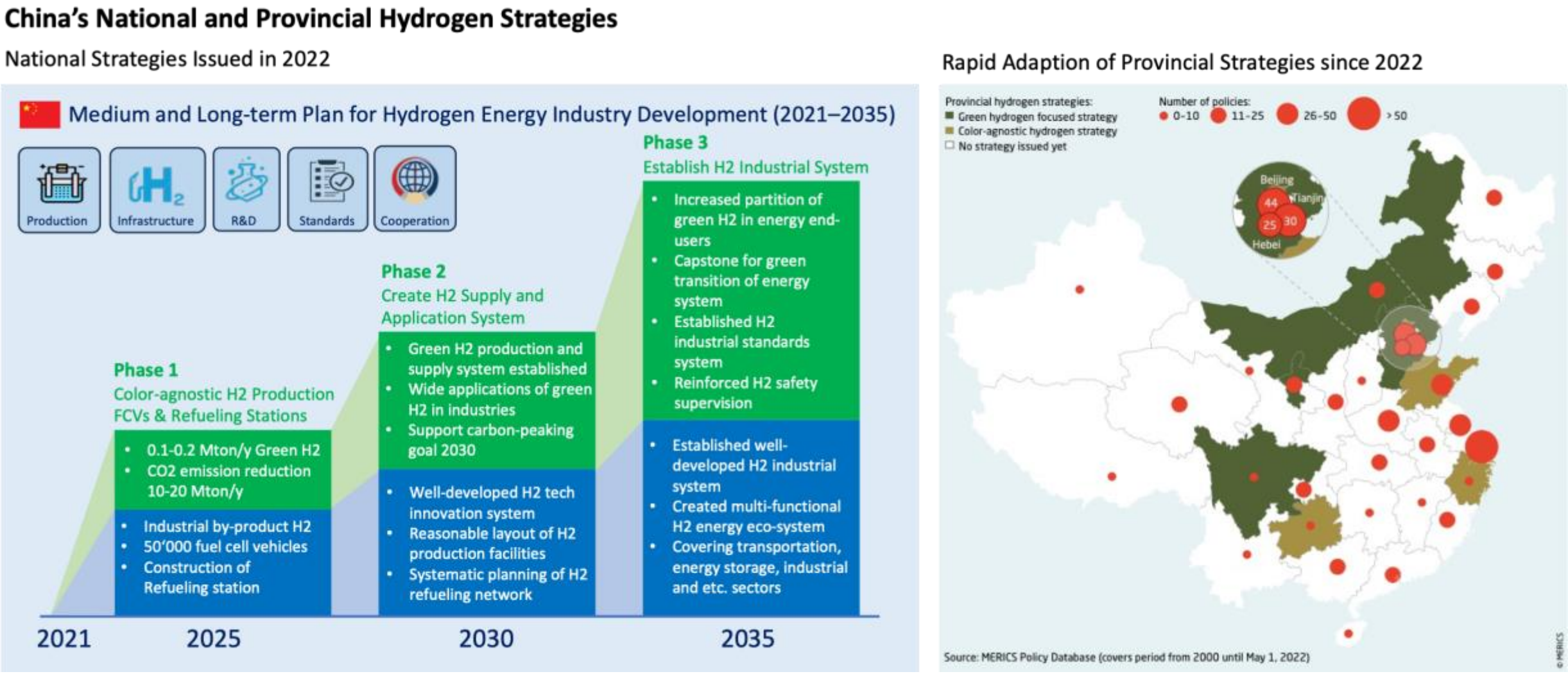

Figure 6: China's national and provincial hydrogen strategies (2021-2035) [15]

### 3.1.2. *State-owned enterprises leading the large-scale and demonstration projects in hydrogen production and infrastructure*

At the upstream of the hydrogen value chain, China's approach to firstly establishing production facilities and infrastructures is leading by state-owned enterprises (SOEs). This is instrumental in overcoming the barriers of large scale investments and complex land use permissions. Pioneering SOEs such as SinoPec, PetroChina, CHN Energy, State Grid, and China Huaneng Group are developing ambitious DEMO and large scale production plans of RH2 and its derivatives in provinces that are rich in renewable energy sources, such as Xinjiang, Inner Mongolia, Ningxia, and Hebei province (Fig.7).

The transition from DEMO projects to scaling up applications is a common strategy of Chinese enterprises for rolling out large scale plans in terms of policy support and funding. By piloting DEMO projects, it allows these SOEs to carefully monitoring the processes and assessing the profitability before widely applied to other regions.

### 3.1.3. *Hydrogen mobility is shaping the demand from end-use sectors*

Navigating to the downstream of the hydrogen value chain, the scene shifts to diversified medium to small-sized private companies in the field of hydrogen applications across various sectors, including the electricity, heating, and mobility. In 2021, more than 70% of the private investments are in the area of fuel cell and hydrogen mobility, according to BCG and Ouyang Minggao CAS Fellow Research Group[18]. These companies are largely located in industrial agglomerate areas like Jiangsu, Zhejiang province in the Yangtze River Delta. Their business is focus on hydrogen utilization and mobility particularly due to the rapidly emerging fuel cell vehicle industry encouraged in national and regional policies. They complement the SOEs' work by addressing the end-use aspects of hydrogen economy, innovating in areas such as hydrogen storage, fuel cell technology, decentralised hydrogen power plans, hydrogen-coupled electric grid and etc. The joint efforts by SOEs and private business crossing the hydrogen value chain, has enhanced the industrial synergy and driven the hydrogen economy forward in China.

## 3.2. *EU Hydrogen Strategy: Green Scaling and Supply Derisking*

### 3.2.1. *EU's hydrogen strategy followed with rapidly issued member countries' strategies*

The EU Hydrogen Strategy was launched in 2020 despite COVID-19 pandemic's socio-economic disruptions, in order to

[17] ChuangYeBang Research Center (2023), China Hydrogen Energy Industry Research Report 2023 (in Chinese), https://m.cyzone.cn/article/736747.html

[18] BCG / Ouyang Minggao Research Group (Aug 2023), Prospects for China's Hydrogen Energy Industry (in Chinese), https://web-assets.bcg.com/a4/e9/7bdf588340169730c3dad382b4b2/bcg-china-hydrogen-industry-outlook-cn-aug-2023.pdf

| Enterprise Name | Logo | Key Hydrogen Projects | Remarks |
|---|---|---|---|
| SPIC (State Power Investment Corporation Limited) | 国家电投 SPIC | • 250Nm2/h proton exchange membrane electrolysis for oxygen production from water.<br>• The first hydrogen refueling station project in **Xiong'an New Area**.<br>• Demonstration of natural gas blending with hydrogen for household use.<br>• **Jilin Da'an** Yiguang integrated demonstration project for green hydrogen and ammonia synthesis. | Electrolysis; H2 refueling; H2 in NG pipe; NH3 |
| SINOPEC (China Petroleum & Chemical Corporation) | 中国石化 SINOPEC | • By 2022, a total of 98 refueling stations had been built, with a combined hydrogenation capacity of approximately 45 tons per day, making it the company with the most refueling stations worldwide.<br>• **Xinjiang Kuche** Green H2 Demonstration Project.<br>• **Inner Mongolia Ordos City** Wind-Solar Integration Green Hydrogen Demonstration Project. | H2 refueling; Green H2; Wind-PV Integration |
| PetroChina (PetroChina Company Limited) | 中国石油 | • Preparation of Proton Exchange Membrane (PEM) Electrolysis Hydrogen Catalysts, and successful batch production of PEM electrolysis hydrogen catalysts.<br>• Breakthrough in the technology for long-distance hydrogen transportation using existing natural gas pipelines.<br>• **PetroChina Sichuan** Petrochemical achieved integrated production, transportation, and marketing in the oil and gas industry. | Electrolysis; H2 in NG pipe; |
| CHN Energy (CHN ENERGY Investment Group Co. LTD) | 国家能源集团 CHN ENERGY | • **Ningxia Ningdong** Renewable Hydrogen Carbon Reduction Demonstration Zone's 620,000 kW Photovoltaic Project.<br>• Wind-Solar Hydrogen-Ammonia Integrated New Energy Storage Demonstration Project.<br>• Guohua (Cangzhou) Comprehensive Energy Co., Ltd.'s 100,000 tons/year synthetic ammonia and supporting project.<br>• Developed 35/70 MPa rapid hydrogenation equipment and large-capacity, low-energy consumption hydrogenation station process control systems, which have been commercially applied in hydrogenation stations. | Green H2; NH3; Electrolysis; |
| State Grid (State Grid Corporation of China) | 国家电网公司 STATE GRID | • **Zhejiang Cixi** Hydrogen-Electric Coupling DC Microgrid Demonstration Project.<br>• **Zhejiang Taizhou** Dachen Island Hydrogen Comprehensive Utilization Demonstration Project.<br>• **Zhejiang Lishui** Xiyun Water-Light-Hydrogen Biomass Nearly Zero Carbon Demonstration Project.<br>• **Zhejiang Hangzhou** Asian Games Low-Carbon Hydrogen-Electric Coupling Application Demonstration Project.<br>• **Anhui Lu'an** Mega-Level Hydrogen Comprehensive Utilization Demonstration Project. | H2-Electric Coupling Grid; |
| CSSC (China State Shipbuilding Corporation) | CSSC 中国船舶集团有限公司 | • **Hebei** Hongmong New Energy Project: 1000Nm³/h Hydrogen Production Equipment.<br>• 500MW Wind Power Hydrogen and Ammonia Integrated Project. | H2 production; NH3 |
| DEC (Dongfang Electric) | DEC 东方电气 | • **Guangzhou Nansha** Electric-Hydrogen Smart Energy Station's Solid-State Hydrogen Energy Generation Project is the first time in China that photovoltaic power generation has been coupled with solid-state hydrogen storage and applied to the power system. | Coupling Grid; H2 (PV/Wind) |
| China Huaneng Group | CHINA HUANENG | • China Huaneng Group's first set of 1300Nm3/h electrolyzers is online, and it is the world's first set of 1300Nm3/h high current density pressure-type alkaline electrolyzers.<br>• Huaneng **Sichuan Pengzhou** 13MW Hydrogen Production from Water Electrolysis Technological Innovation Demonstration Project."<br>• **Inner Mongolia Chifeng** Wind-Solar High Proportion Blending Green Hydrogen Production Demonstration Project." | AE Electrolysis; H2 (PV/Wind) |

Data source: Chuangyebang Research Center

Figure 7: State-owned enterprises leading hydrogen production projects in China (2023) [17]

implement the ambition of the European Green Deal for achieving the first climate-neutral continent by 2050[19]. The plan sets a vision that EU can turn clean hydrogen into a viable solution to decarbonize industrial sectors, including the hard-to-abate ones. It also unfolds in three strategic phases (Fig.8) that identify the challenges to overcome, lays out the levers that the EU can mobilise and presents a roadmap of actions for the coming years[20].

The phase 1 (2020-2024) focuses on establishing a robust regulatory framework for a competitive hydrogen market and incentivizing the color transition from blue/brown hydrogen to the specifically targeted RH2. The EU aims to have at least 6 GW of electrolyser capacity by 2024, producing up to 1 Mt of RH2, targeting initial decarbonization of existing hydrogen production and its integration into industrial processes and heavy-duty transport. It also includes the transformation of some fossil-based hydrogen production facilities with carbon capture and storage (CCS) technology.

The Phase 2 (2025-2030) is expected to boost RH2 as cost-competitive compared with other forms of hydrogen production, via dedicated demand side policies for new applications adoptions in industry. This phase is set to expand electrolyser capacity to 40 GW, producing up to 10 Mt of RH2, with the establishment of a network of hydrogen refuelling stations and large-scale storage. Local hydrogen clusters, or known as "Hydrogen Valleys", will develop relying on local production and demand of hydrogen. This phase also envisions planning a basic EU-wide hydrogen grid by adapting parts of the existing gas infrastructure.

The Phase 3 (2030-2050) anticipates hydrogen technologies reaching maturity, permeating all sectors that are hard-to-abate, including aviation, maritime, and industrial buildings. A significant increase in renewable energy production will be necessary, as RH2 could account for up to a quarter of the EU's renewable electricity by 2050. The three-phase goals assured the EU's commitment to decarbonization for climate neutrality and emphasized the important role of hydrogen in enhancing energy security during the green transition.

The EU-level Hydrogen Strategy has been complemented by rapidly released member countries' national strategies as shown in Fig.8. In 2020, 6 EU countries issued national hydrogen strategy and 7 countries had their strategies in development. The sum of planned electrolyser capacity from 7 countries (Austria, France, Germany, Italy, Netherlands, Portugal, Sweden) is 27.8 GW by 2030, which already accounts for 70% of the EU 2030 target of 40 GW.

Energy security has emerged as a critical concern for the EU due to the energy market disruption triggered by the escalation of Russo-Ukrainian War since Feb 2022. In response to the domestic hardships, EU launched the REPowerEU plan in May 2022, which aimed to diversify and de-risk the EU energy supply, and specifically secured the strategic partnerships with Namibia, Egypt and Kazakhstan to ensure a secure and sustainable supply of renewable hydrogen[22].

### 3.2.2. *Boosting demand and scaling up hydrogen production in the EU*

The market price of RH2 is subjected to the economies of scale and technology maturity. The Key actions of EU hydrogen strategy focus on boosting demand and scaling up production for hydrogen economy. The strategy proposed measures to facilitate the use of hydrogen and its derivatives in the transport sector, according to the EU Sustainable and Smart Mobility Strategy published in 2020[23]. The strategy emphasizes the importance of RH2, supported by demand-side policies in end-use sectors. The proposal for the Renewable Energy Directive's revisions, including specific sub-targets for renewable hydrogen in industry and transport, highlighted this green focus as well.

[19] European Commission (2020), The European Green Deal, https://commission.europa.eu/strategy-and-policy/priorities-2019-2024/european-green-deal_en

[20] European Commission (2020), A hydrogen strategy for a climate-neutral Europe, https://eur-lex.europa.eu/legal-content/EN/TXT/?uri=CELEX:52020DC0301

[21] Hydrogen Europe (2023), Map of EU countries that adopted hydrogen strategy and in development phase, https://hydrogeneurope.eu/sites/default/files/Map_%20National%20H2%20Strategies.pdf

[22] European Commission (2022), Communication REPowerEU Plan COM(2022)230 (europa.eu), https://commission.europa.eu/publications/key-documents-repowereu_en

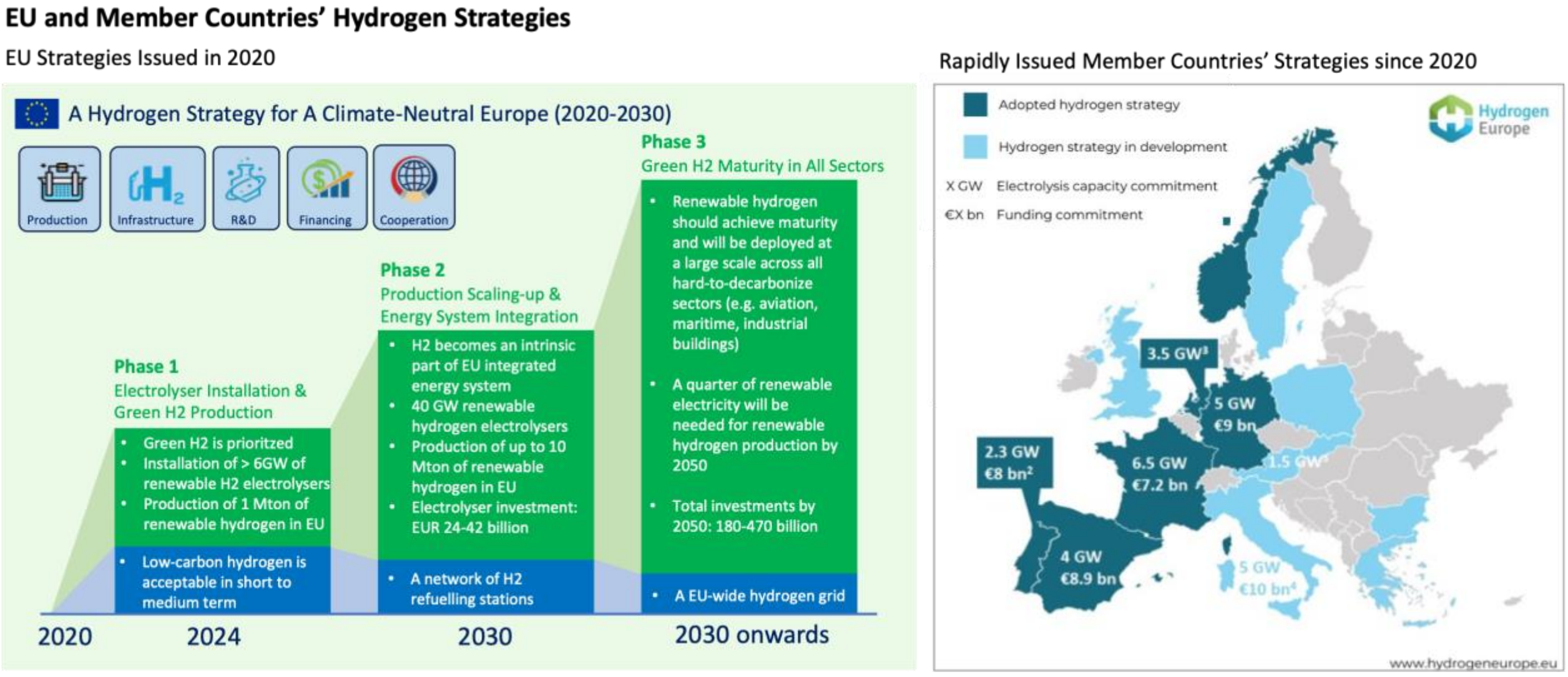

Figure 8: EU hydrogen strategy and the map of member countries' strategies (2020) [21]

In parallel, the EU is working on introducing a common low-carbon threshold/standard for the promotion of hydrogen production installations, and developing comprehensive certification criteria for renewable and low-carbon hydrogen. Another innovative aspect of the strategy is the development of a pilot scheme for a "Carbon Contracts for Difference" programme, supporting the production of low-carbon and circular steel and basic chemicals. Together with an enabling and supportive framework including the proposal for Trans-European Energy Infrastructure, Alternative Fuels Infrastructure, Trans-European Transport Network, Clean Hydrogen Partnership, and European Hydrogen Valley partnerships, these key actions represent comprehensive and multiple-faceted approaches to achieve EU's ambitious climate goals.

### *3.2.3. European Hydrogen Bank is on agenda alongside with InvestEU programme*

Strategic investments in green hydrogen are in the window of the EU Recovery Plan and InvestEU programme, which supports the REPowerEU plan with ambitious and targeted investment figures. In 2020, the European Commission estimated requirements of €24 - 42 billion for electrolysers and €220 – 340 billion to scale up 80 – 120 GW of solar and wind energy capacity for these electrolysers. Moreover, €11 billion is allocated to renovate half of the plants producing fossil-based hydrogen with CCS technologies, and €65 billion is dedicated to hydrogen transport, distribution, and storage infrastructure, including refuelling stations[24]. The established European Clean Hydrogen Alliance in 2021 further strengthens this investment agenda, facilitating a coordinated investment flow between public authorities and industry[25]. The EU Recovery and Resilience Scoreboard shows that at least €9.3 billion were allocated for hydrogen projects in the recovery plans of 15 EU countries in 2021[26].

The world first European Hydrogen Bank (EHB) was announced in September 2022 as a significant establishment from the EU, representing a strategic financial institution within the InvestEU programme (Fig.9). This hydrogen-specific pioneering bank aims to support the scale-up of domestic hydrogen production and secure diversified hydrogen imports from outside the EU. The Bank receives funding from EU Innovation Fund for domestic supply-side auctions allocating fixed premium payments to hydrogen producers within the EU, and multiple other funds for international producers [28]. The first auction under the Bank was announced in November 2023 with an initial €800 million to support the EU production of renewable hydrogen. The fixed premium intended to bridge the current cost disparity between producing RH2 and the price consumers willing to pay. Through the pilot auction, it outlines the max-imum support level of 4.5 €/kg renewable hydrogen [29], which is competing with the estimation of 3-5 €/kg for hydrogen imports by ship [30]. In the long rum, the Bank is expected to unlock private investments inside and outside the EU by addressing the initial investment challenges and needs, bridging the investment gap and connecting future renewable hydrogen supply to EU consumers.

[23]European Commission (Dec 2020), Sustainable and Smart Mobility Strategy – putting European transport on track for the future, https://eur-lex.europa.eu/legal-content/EN/TXT/?uri=CELEX%3A52020DC0789

[24]European Union (2021), InvestEU Programme, https://investeu.europa.eu/investeu-programme_en

[25]European Commission (Jul 2020), European Clean Hydrogen Alliance, https://single-market-economy.ec.europa.eu/industry/strategy/industrial-alliances/european-clean-hydrogen-alliance_en

[26]EU Commission (Dec 2021), Recovery and Resilience Scoreboard, https://ec.europa.eu/economy_finance/recovery-and-resilience-scoreboard/assets/thematic_analysis/1_Clean.pdf

[27]European Hydrogen Bank Schematic (European Commission) and InvestEU Programme (European Clean Hydrogen Alliance)

[28]European Commission (2023), 2023 pilot auction to produce renewable hydrogen, https://climate.ec.europa.eu/eu-action/eu-funding-climate-action/innovation-fund/competitive-bidding_en#ref-2023-pilot-auction-to-produce-renewable-hydrogen

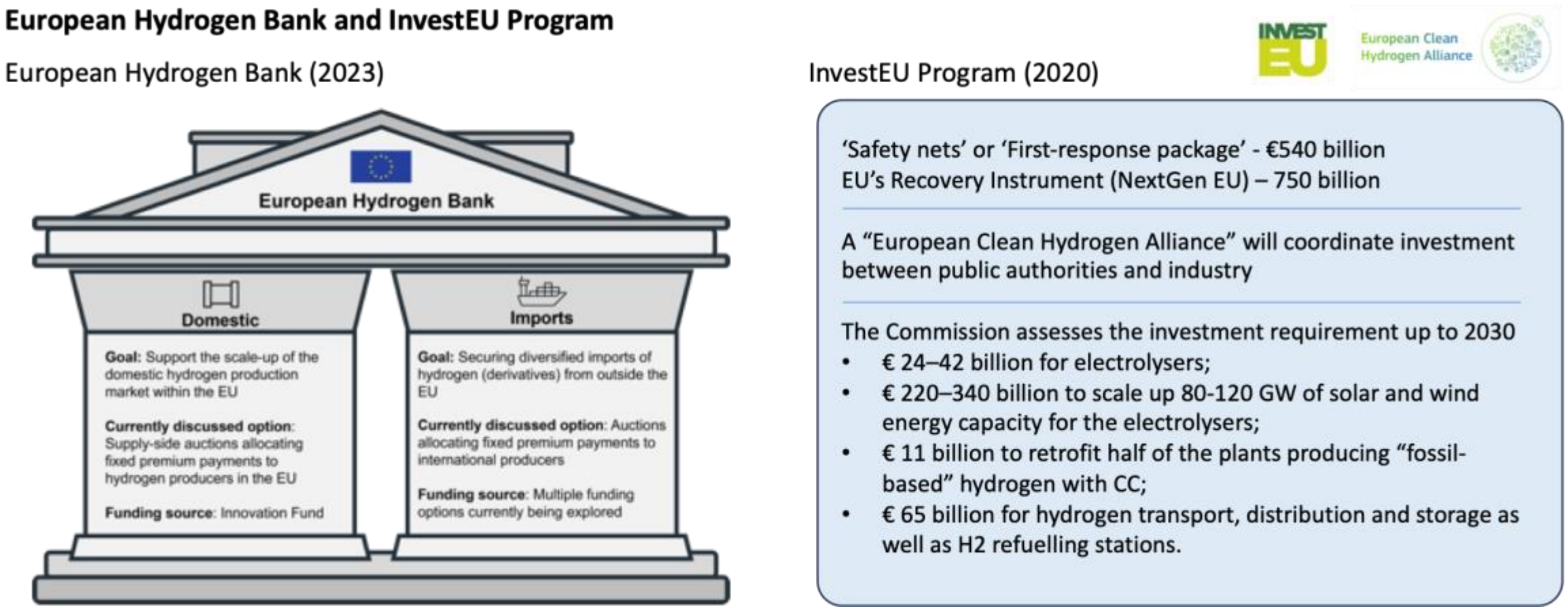

Figure 9: Schematic of European Hydrogen Bank (2023) and InvestEU Programme (2021) [27]

### 3.3. *Comparative analysis of strategic hydrogen development in China and EU*

#### 3.3.1. *China is leading the core hydrogen infrastructure developments*

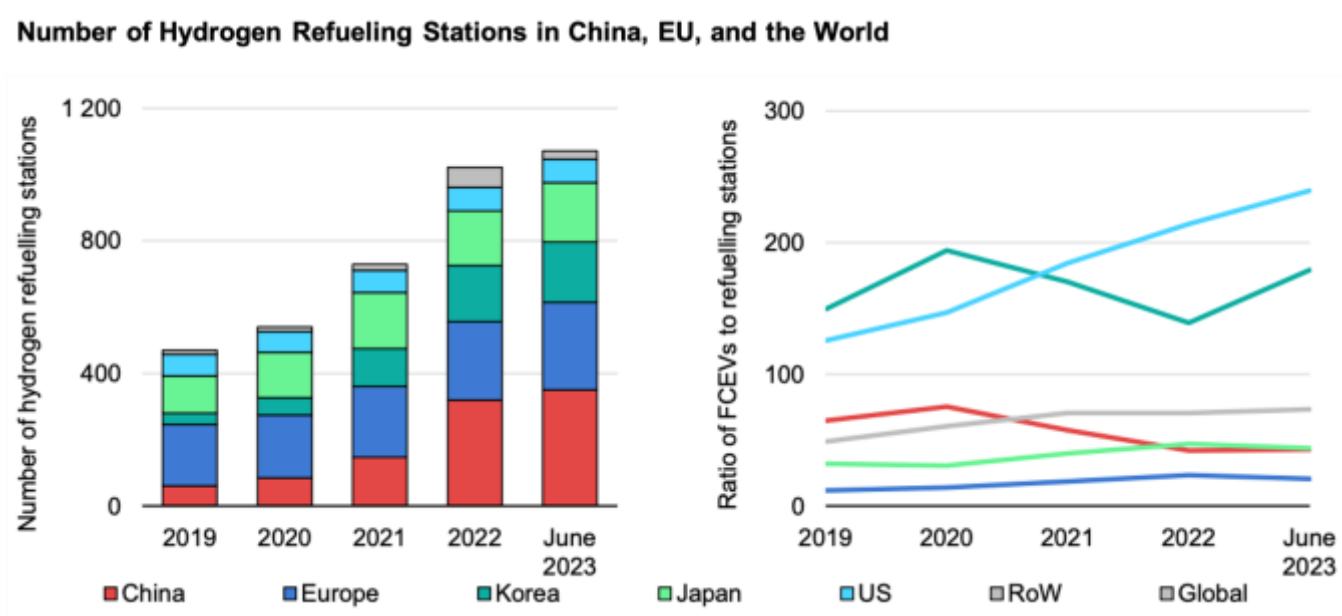

Figure 10: Statistics of hydrogen refuelling stations in China, EU, and the world (2023) [32]

Hydrogen infrastructures have been integral to lay the foundation of the hydrogen economy, which is often described as the "Chicken-and-Egg Dilemma". At the core of the infrastructures are hydrogen refuelling stations that support end-user markets such as hydrogen powered vehicles and relevant industries.

In China, the number of hydrogen refuelling stations has surpassed 358 until the end of 2022[33], accounting for more than 30% of the world statistics as shown in Fig.10. Geographically, these stations are located mostly in three major economic areas: the Capital Economic Circle (CEC), the Yangtze River Delta (YRD), and the Pearl River Delta (PRD) Region as shown in Fig.11. This distribution indicates the hydrogen activities driven by local policy and industries in react to the local hydrogen demand and production.

Similarly, in the EU, the hydrogen refuelling infrastructure is growing at a pace comparable to China. However, the distribution of stations across Europe is more widespread across EU member countries. Few concentrated areas are identified in the Benelux area (Belgium, Netherlands, Luxembourg), Northwestern France, and Western Germany. In particular, Germany stands out with the highest number of refuelling stations in Europe, highlighting its leading role in the EU hydrogen economy, and aligning with its ambitious National Hydrogen Strategy updated in 2023[34].

#### 3.3.2. *EU's diversified electricity sources for hydrogen production compared to China*

In China, the landscape of hydrogen production is shaped by the types of electricity sources used, indicating by the color of hydrogen with different level of GHG emissions. Based on the IEA Hydrogen Database published in October 2023 as shown in Fig.12, the current predominant sources for RH2 production are wind and solar energy, including both onshore and offshore wind. The production of yellow hydrogen (derived from the grid) and pink hydrogen (generated through nuclear energy) are considered minor. Future unveiled productions indicate that there is an increasing trend of RH2 production utilising various renewables in China.

[29] European Commission (Nov 2023), Commission launches first European Hydrogen Bank auction with €800 million of subsidies for renewable hydrogen production, https://ec.europa.eu/commission/presscorner/detail/en/IP_23_5982

[30] European Commission (Mar 2023), Communication From The Commission To The European Parliament, The Council, The European Economic And Social Committee And The Committee Of The Regions on the European Hydrogen Bank, https://energy.ec.europa.eu/system/files/2023-03/COM_2023_156_1_EN_ACT_part1_v6.pdf

[32] IEA (2023), Global Hydrogen Review 2023, https://iea.blob.core.windows.net/assets/ecdfc3bb-d212-4a4c-9ff7-6ce5b1e19cef/GlobalHydrogenReview2023.pdf

[33] China Energy News/People.cn (Jul 2023), Why are hydrogen refuelling stations "not full"? (in Chinese), http://paper.people.com.cn/zgnyb/html/2023-07/17/content_26007345.htm

[34] BMWK (Jul 2023), German National Hydrogen Strategy Update, https://www.bmwk.de/Redaktion/EN/Publikationen/Energie/national-hydrogen-strategy-update.pdf?__blob=publicationFile&v=2

[35] H2stations.org by LBST (2023), Distribution of Hydrogen refuelling Stations in China and Europe, https://www.h2stations.org/press-release-2023-another-record-addition-of-european-hydrogen-refuelling-stations-in-2022/

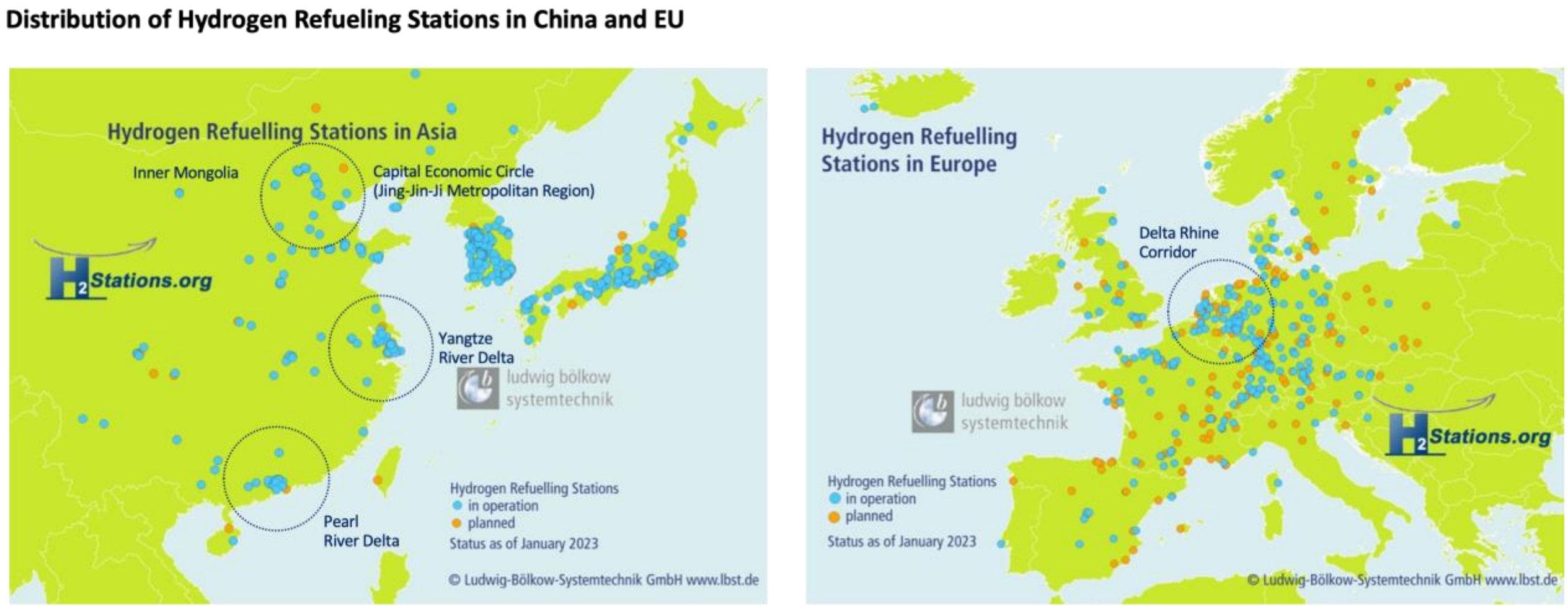

Figure 11: Distribution of hydrogen refuelling stations in China and Europe (2023) [35]

**Hydrogen Production by Electricity Source in China and EU**

Unit: IEA Estimated Normalised Capacity [Nm³ $H_2$/hour]

| | | China (Current) | China (Future) | EU27 (Current) | EU27 (Future) | GHG |
|---|---|---|---|---|---|---|
| Green | Solar PV | 201.5K | 33.09K | 32.81K | 7.32M | Minimal |
| | Onshore wind | 375K | 7.76K | 51.76K | 271.57K | |
| | Offshore wind | 233.21K | 0 | 160.74K | 13.05M | |
| | Hydropower | 6K | 0 | 175.37K | 192.27K | |
| | Various Renewables | 668.79K | 1.98M | 35.37K | 5.28M | |
| Yellow | Grid+Renewables | 0 | 0 | 39.47 | 416.67K | Medium |
| | Grid | 192.31 | 0 | 179.02K | 2.69M | |
| Pink | Nuclear | 50.1 | 50K | 152.17 | 41.18K | Minimal |
| | Other/Unknown | 98.65K | 390.72K | 180.59K | 10.28M | |

Figure 12: Hydrogen production by electricity sources in China and the EU (2023) [37]

In contrast, the EU's current RH2 production powered by diversified renewable electricity sources including solar, wind, and hydropower. Compared with China's reliance on wind and solar production, EU extends the portfolio particularly with considerable scale of hydropower and grid electricity. In the future, EU plans to largely increase the offshore wind based RH2 production, as well as the solar based production. This landscape reflects EU's strong adaptability according to its natural resources and existing infrastructures, towards scaling up hydrogen production and reducing carbon emissions.

[37]IEA (Oct 2023), Hydrogen Production and Infrastructure Projects Database, Electricity Source in China and the EU, https://www.iea.org/data-and-statistics/data-product/hydrogen-production-and-infrastructure-projects-database

### *3.3.3. Emerging and conventional end-use sectors for hydrogen projects in the EU and China*

Towards the end-use sectors of hydrogen value chain, the statistics of hydrogen projects from IEA database also show distinctive focus in China and the EU (Fig.13). The end-use sectors for hydrogen projects in China are leading by the mobility (27%) and ammonia (23%), representing half of the total hydrogen projects in China. In addition, the methanol sector, contributing 12%, and the Refining sector, accounting for 11%, also hold significant shares in the hydrogen projects. This indicates the demand of hydrogen in China arises from mainly the traditional ammonia, methanol, refining sectors beyond the emerging mobility sector.

The EU explores diversified applications of hydrogen across different sectors compared to China. Mobility remains a major part of the hydrogen projects with a 29% share. Notably, the hydrogen applications in grid injection sectors account for 9% share in the EU, which is not explored by China's projects yet. The Ammonia sector, which is significant in China, holds a minor position in the EU with only 3.9% of the projects. In Germany, the distribution of hydrogen projects across end-user sectors generally aligns with the major trends observed in the EU statistics, with slightly increasing shares in power, iron and steel sectors.

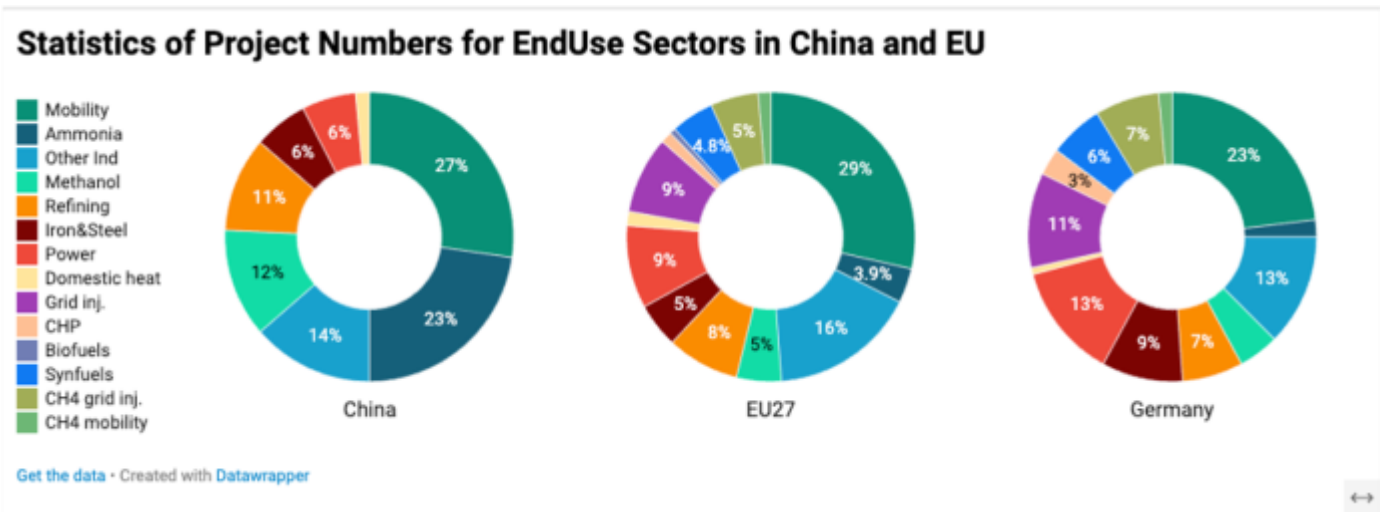

Figure 13: Hydrogen projects for various end-use sectors in China and the EU (2023) [39]

[39]IEA (Oct 2023), Hydrogen Production and Infrastructure Projects

### *3.3.4. Divergent electrolyser technology roadmaps in China and the EU*

Electrolyser serves as the primary technology for RH2 production, thus the technological roadmap adopted by China and the EU can shape the hydrogen economy in the long term. The Alkaline electrolyser (AE or ALK) was the world first hydrogen production electrolyser that are fully matured in the market with the lowest CAPEX among all 4 types of electrolysers as shown in Fig.15. The Proton Exchange Membrane (PEM) Electrolyser is considered as a nearly-matured and competitive technique for high purity and efficient hydrogen production compared with AE (ALK) technique. The Anion Exchange Membrane (AEM) Electrolyser and Solid Oxide Electrolyser Cell (SOEC) are highly attractive for scientific research projects, however, not yet market-ready with a relatively lower Technology Readiness Level (TRL).

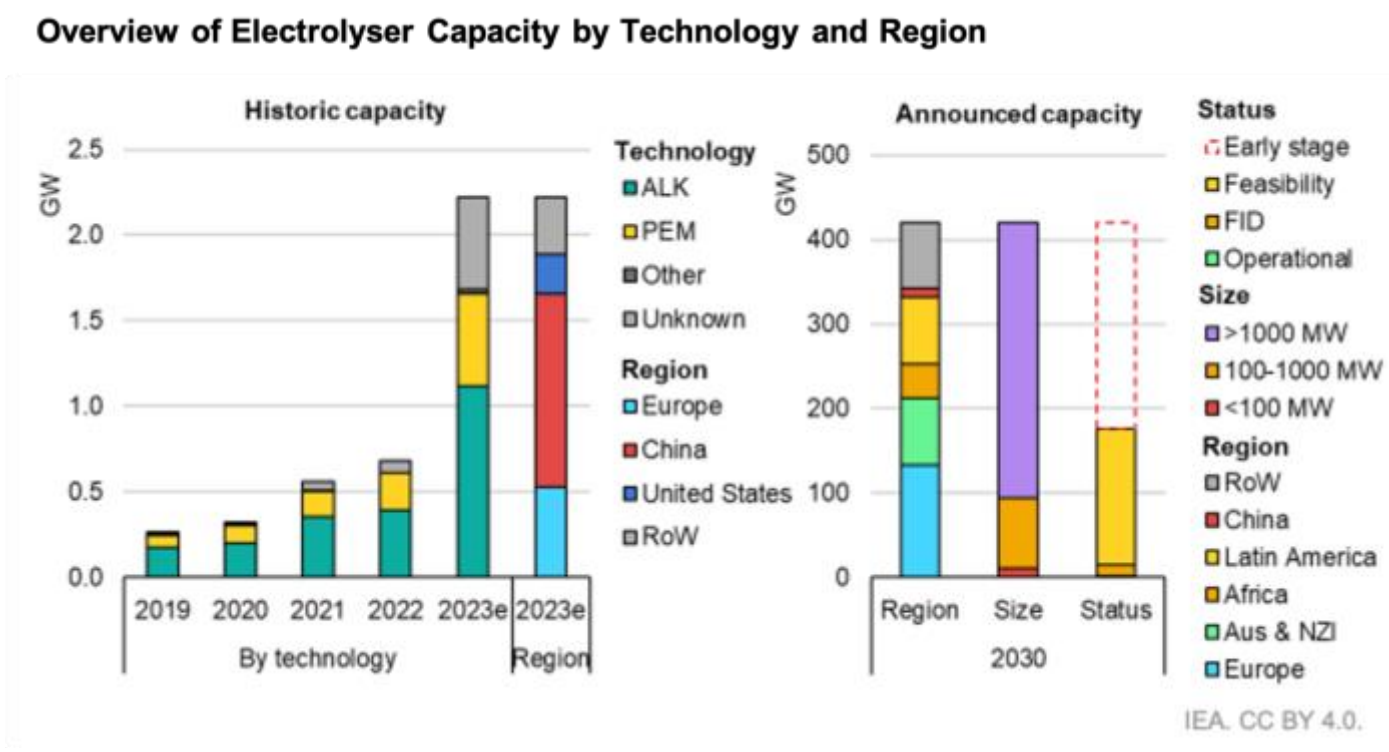

Figure 14: Statistics of electrolyser capacity by technology and region (2023) [41]

According to IEA statistics in Global Hydrogen Review 2023 (Fig.14), Alkaline Electrolyser accounts for almost half the estimated capacity in 2023, with a share of roughly 25% for PEM electrolyser. This clearly shows the different market penetration levels of AE and PEM electrolysers. Categorized by region, China is leading the electrolyser capacity with more than 50% share (AE type dominated), while Europe's share is nearly 25% (PEM type dominated). Towards the Phase 2 (2025 - 2030) outlined in EU Hydrogen Strategy, hydrogen projects announced in EU aims to achieve more than 120 GW capacity which accounts for nearly 30% of the world capacity. Whereas in China's National Hydrogen Strategy, there was no specific capacity goals announced for comparison. However, single projects leading by State-owned Enterprise in China can be significant and are expected to be rapidly scaled up after successful demonstrations. For instance, a 260 MW plant by Sinopec have started operations in July 2023, and more than 550 MW of projects are under construction.

## 4. Case Study: Contrasting Hydrogen Economic Areas in China and Europe

### *4.1. Northern China - Inner Mongolia and Capital Economic Circle (CEC)*

#### *4.1.1. Inner Mongolia: Resource-driven integrated renewables for RH2*

Inner Mongolia, as China's largest renewable production region and the second largest coal production region, has been intensively invested by the state-owned enterprises in China through commissioning large hydrogen demonstration projects (Fig.16). The regional government rolled out the "14th Five-Year Plan" (2021-2025) for hydrogen development in February 2022, envisioning a hydrogen industry worth over RMB 100 billion (€12.7 billion) towards 2025[43]. With an advantageous low cost of renewable power in Inner Mongolia, the plan aims to carry out more than 15 DEMO projects of types such as 'multi-energy complementary + hydrogen' and 'power-grid-load-storage integration + hydrogen', supplying 1.6 Mt H2/y by 2025 with a share of RH2 more than 30%. The plan also specified the goal for 60 hydrogen refuelling stations (3 as of 2020) and 5,000 fuel cell vehicles, introducing more than 50 enterprises across the hydrogen industry chain by 2025.

Statistics by GaoGong Industrial Institute (GGII) show that there are at least 40 ongoing RH2 projects in Inner Mongolia at various launching phases, with a total investment exceeding RMB 250 billion (about €31.95 billion) as of July 2023[44]. 22 out of 40 projects have announced their expected operational dates, ranging from 2023 to 2028. State-owned enterprises (SOEs) are the major responsible entities behind the projects, including Sinopec, PetroChina, China Energy Group, China General Nuclear Power Group, China Coal, Huaneng, and etc.

Leveraging the abundant renewables, Sinopec launched the World's Largest Green Hydrogen-Coal Chemical Project in Erdos, Inner Mongolia in February 2023[45]. This project targets production capacity of 30,000 tons H2/y, which will be used for the carbon reduction initiatives of the adjacent ZTHC Energy intensive coal processing pilot project in Erdos. Backed by an investment of RMB 5.7 billion (about €728.4 million) for this project, Sinopec is expected to cut carbon emissions by 1.43 Mt/y, moreover, contributing RMB 600 million (about €76.6 million) to regional GDP and RMB 30 million (about €3.8 million) to tax revenue. Undoubtedly, it is a new milestone after Sinopec's large-scale green hydrogen pilot project

Database, Hydrogen projects for vairous end-use sectors in China and the EU, https://www.iea.org/data-and-statistics/data-product/hydrogen-production-and-infrastructure-projects-database

[41] IEA (2023), Global Hydrogen Review 2023, https://iea.blob.core.windows.net/assets/ecdfc3bb-d212-4a4c-9ff7-6ce5b1e19cef/GlobalHydrogenReview2023.pdf

[42] Information Source: sbh4 consulting for figures; IRENA for the pros and cons of different elctrolyser technologies; DOE for the Technology Readiness Level Assessments

[43] Inner Mongolia Energy Bureau (Feb 2022), Inner Mongolia Autonomous Region's "14th Five-Year" Hydrogen Development Plan (2021-2025) (in Chinese), https://www.nmgxny.com/myloads/soft/220301/1-220301lT002.pdf

[44] GaoGong Industrial Institute (GGII) (Jul 2023), Overview of 40 RH2 projects in Inner Mongolia (in Chinese), https://www.escn.com.cn/20230730/b1485dee2f334cb39fb5cf9b474e663e/c.html

[45] Sinopec (Feb 2023), Sinopec launches the World's Largest Green Hydrogen-Coal Chemical Project in Inner Mongolia, http://www.sinopecgroup.com/group/en/Sinopecnews/20230331/news_20230331_502557647316.shtml

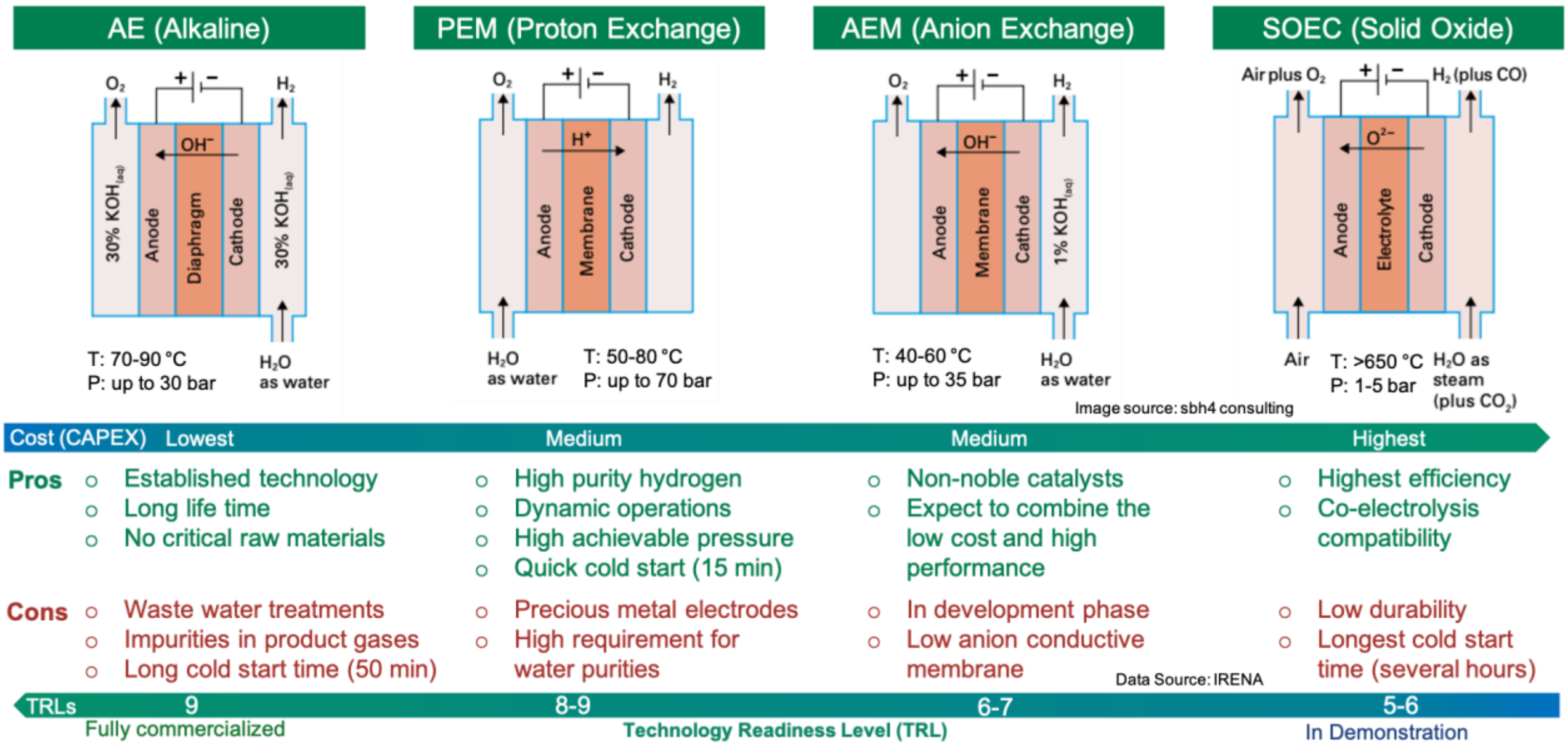

Figure 15: Overview of electrolyser technologies with corresponding CAPEX and TRLs (2023) [42]

in Kuqa, Xinjiang in 2021[46].

In parallel, Sinopec is also planning a "Grid-coupled Type of Wind-Solar Integrated Hydrogen Demonstration Project" in Ulanqab, Inner Mongolia [47]. This DEMO project aims to produce 100,000 tons H2/y with a planned total investment of RMB 20.5 billion (about €2.87 billion). The scale of the renewable installation is in total of 2,546 MW, including 1,742 MW of wind power and 804 MW of solar power. The construction is scheduled to start in December 2023, and the project is expected to be operational by June 2027.

Inner Mongolia, as a successful model for resource-driven integrated renewables, has been remarked by world-class hydrogen production projects as illustrated above. SOEs such as Sinopec launched ambitious hydrogen production projects rapidly, following the regional government's hydrogen development plan in 2022. The number of ongoing projects and the total investments have already exceeded the government's conservative goals in hydrogen development plan, signaling the strong industrial capability and synergy in navigating the regional hydrogen economy.

### 4.1.2. *Capital Economic Circle: Load center for hydrogen mobility and green industry*

Beijing, the core of Capital Economic Circle (CEC), is the first city to host both Summer and Winter Olympic Games in 2008 and 2022, respectively. The CEC is also known as Jing-Jin-Ji Metropolitan Region that aims to integrate Beijing, Tianjin and Hebei Province with a total population share of 7% and GDP share of 8.29% in 2022[49]. Hydrogen powered vehicles were highlighted during both Olympic games: 20 fuel cell cars were employed in 2008[50] and more than 1000 fuel cell buses and cars were deployed in 2022[51]. The increasing deployments of hydrogen mobility in large events in CEC, not only show the regional interests and technological advancements, but also the calling for green mobility and industry in CEC —- the load center of northern China.

All three governments in CEC announced their hydrogen specific plans since 2020 with common interests in carbon emission reduction, hydrogen refuelling station constructions, and hydrogen fuel cell vehicle (FCV) developments. Beijing's hydrogen industry development plan (2021-2025) outlined the overall goal of CEC to reach a hydrogen industry scale of RMB 50 billion (about €6.39 billion) and carbon emission reduction of 1 Mt by 2023. Both numbers were projected to be doubled by 2025[52]. The hydrogen plans also specified the goals for 37 refuelling stations and 10,000 FCVs in Beijing by 2025, 10 re-fuelling stations and 1,000 FCVs in Tianjin by 2022, 100 refuelling stations and 10,000 FCVs in Hebei by 2025. According to the statistics by ChinaH2Data, the numbers of constructed refuelling stations in Beijing, Tianjin, and Hebei are 17, 13, and 33 by the end of 2023[53]. Focusing on the industrialization of FCVs, CEC aimed to collectively establish industry chain of hydrogen production, storage, transportation, refuelling, and usage. The Beijing plan was designed to facilitate complementary advantages, staggered development, and mutual benefits for Tianjin and Hebei Province. These policies indicate the potential increasing demand for RH2 from mobility sector in a fast pace.

[46]Sinopec (Aug 2023), Xinjiang Kuqa Green Hydrogen Demonstration Project has been fully completed and put into operation (in Chinese), http://www.sinopecgroup.com/group/xwzx/gsyw/20230831/news_20230831_309591498891.shtml

[47]Souhu News (Dec 2022), With an investment of 23.5 billion RMB, Sinopec's two major hydrogen energy projects in Inner Mongolia will start construction next year (in Chinese), https://www.sohu.com/a/621866924_157504

[48]Overview of Inner Mongolia and Capital Economic Circle and Hydrogen Pipeline Project (depicted according to public information); Statistics in the table according to https://www.energypolicy.columbia.edu/publications/chinas-hydrogen-strategy-national-vs-regional-plans/

[49]NRDC (Feb 2023), The total GDP of Beijing, Tianjin and Hebei exceeds 10 trillion RMB in 2022 (in Chinese), https://www.ndrc.gov.cn/xwdt/ztzl/jjyxtfz/202302/t20230224_1349473.html

[50]Ministry of Science and Technology of PRC (Jul 2008), New energy vehicles support green Olympics (in Chinese), https://www.most.gov.cn/tpxw/200807/t20080711_63009.html

[51]Xinhua Net (Feb 2022), A detailed breakdown of "CIMC equipment" for the Beijing Winter Olympics (in Chinese), http://www.xinhuanet.com/fortune/2022-02/18/c_1211578226.htm

[52]Beijing Municipal Bureau of Economy and Information Technology (Aug

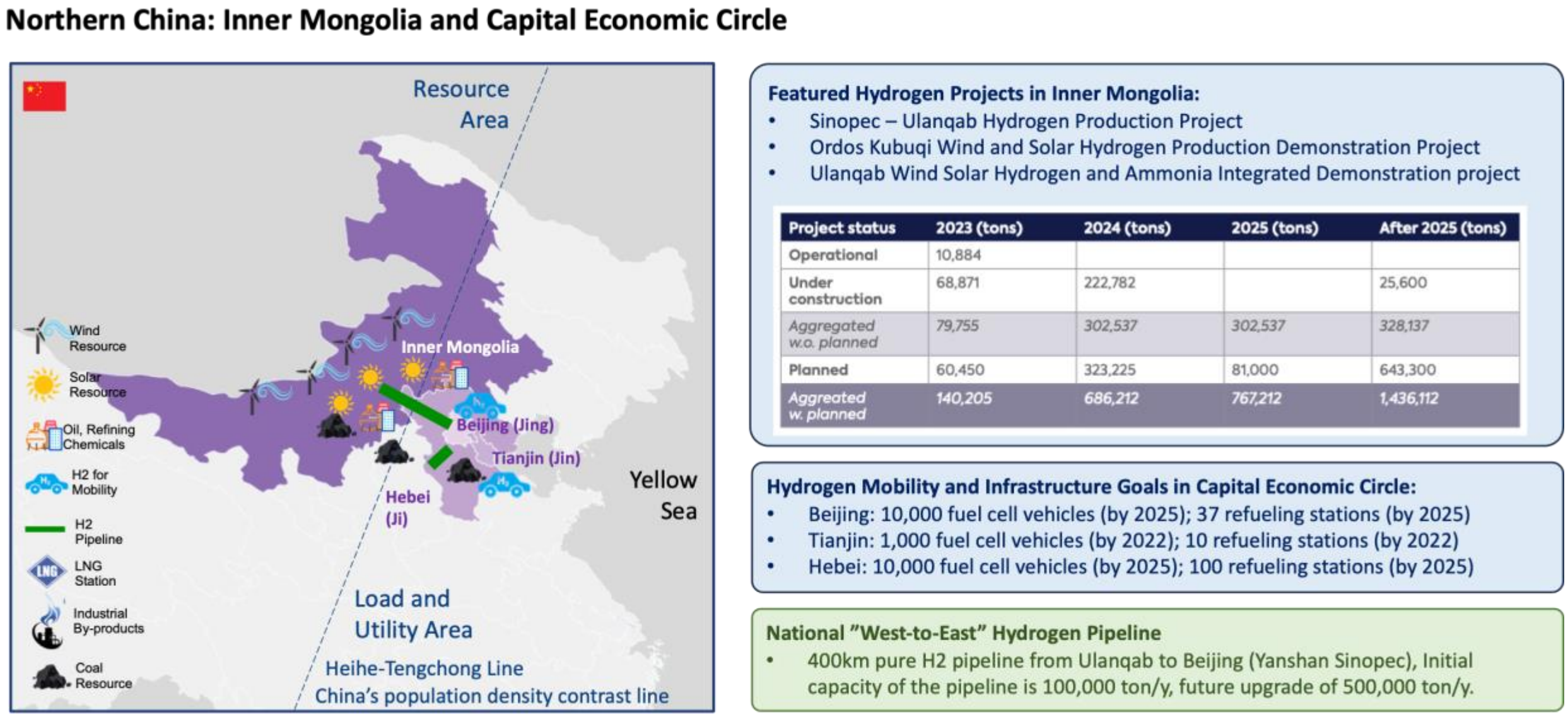

Figure 16: Overview of Inner Mongolia and Capital Economic Circle with "West-to-East" national hydrogen pipeline (2023) [48]

Over-viewing the industrial landscapes of the CEC, Beijing and Tianjin have imposed strict regulations on high-emission industries, resulting the transfer of hard-to-abate sectors to nearby Hebei province with vast plain land. The steel industry is at the core of Hebei's economy. In 2022, the steel industry accounting for 29.8% of the province's total industrial revenue[54]. Moreover, coal accounts for about 66% of Hebei's total energy consumption in 2022[55]. These facts impose significant challenges for Hebei towards "greened" industry and energy transition, calling for RH2 supply to the region urgently.

2021), Beijing Hydrogen Energy Industry Development Implementation Plan (2021-2025) (in Chinese), https://www.ncsti.gov.cn/zcfg/zcwj/202108/t20210816_38829.html

[53]Big Data Platform for Hydrogen Industry (Dec 2023), Statistics of hydrogen refuelling stations (in Chinese), https://www.chinah2data.com/

[54]People's Net (Nov 2023), The "three modernizations" transformation of the steel industry in Hebei (in Chinese), http://he.people.com.cn/n2/2023/1124/c192235-40654017.html

[55]China Electric Council (Oct 2023), Hebei Province is accelerating the construction of a new energy system (in Chinese), https://www.cec.org.cn/detail/index.html?3-326344

#### *4.1.3. National "West-to-East" Hydrogen Pipeline: Supplying RH2 from Inner Mongolia to load center of Capital Economic Circle*

Just as all roads lead to Rome, all pipelines converge towards the metropolitan city of Beijing. Historically in China, the national "West–to-East" Gas Pipeline was put into operation in 2005 and operated by PetroChina. Such national scale project, aimed at mobilizing national resources, signaling China's aspirations for infrastructure developments with long term payback period.

In April 2023, the Sinopec "West–to-East" pure hydrogen pipeline (Fig.16) demonstration project in China was the first time included in the "National Unified Network Construction and Implementation Plan for Petroleum and Natural Gas" [56]. This pipeline starts in Ulanqab Inner Mongolia, and ends in sub-urban of Beijing, covering more than 400 kilometers. It marks as the first cross-provincial, large-scale, long-distance pure hydrogen pipeline in China. Once operational, the pipeline is aiming to replace the fossil fuel-based hydrogen production in CEC.

The pipeline's initial transport capacity is 100,000 tons H2/y, with a potential future expansion to 500,000 tons H2/y. Additionally, multiple exit outlets will be reserved along the pipeline for including potential hydrogen sources. As explained in section 4.1.1, Sinopec already planned 100,000 tons H2/y DEMO projects in Ulanqab that targets to transport via the "West–to-East" hydrogen pipeline to Sinopec branch in Yanshan, Beijing. The Ulanqab-Beijing Pipeline is significant in China's hydrogen development. It is not just the country's first long-distance hydrogen pipeline, but it also serves as a pioneering model for addressing the RH2 supply-demand imbalance in Northern

[56]Xinhua Agency (Apr 2023), First "West-to-East" Hydrogen pipeline is included in the national plan (in Chinese), https://www.gov.cn/yaowen/2023-04/10/content_5750685.htm

China. When considering even longer distance hydrogen transportation, especially with a matured electricity market in China, the economic advantage tilts towards ”ultra-high-voltage transmission + load-side hydrogen production” over the alternative approach of ”source-side electricity for hydrogen production + hydrogen pipeline to load-side.”

### *4.2. Eastern Coastal China - Yangtze River Delta (YRD) Hydrogen Corridor*

#### *4.2.1. One Metropolis: Shanghai off-shore wind hydrogen production and hydrogen seaport development*

Since the Tang Dynasty (618-907 C.E.), the Yangtze River Delta (YRD) has been the leading economy in China due to its rich business culture, strategic river connections, bustling trading ports, throughout China’s history. Nowadays, the YRD region contributes to around 25% of the national GDP[57] with a population share of 16.7%[58]. It hosts the largest industrial agglomerations and ports in the country, making it the primary load center of eastern coastal China.

Prior to the release of ”14th Five-Year” National Hydrogen Plan in 2022, the ”YRD Hydrogen Corridor Construction and Development Plan” (Fig.17) was published early in May 2019 before the COVID-19 pandemic. The plan was supervised by the YRD Regional Cooperation Office and edited by Chinese Society of Automotive Engineers[59]. It focuses on the enabling the interconnections of hydrogen infrastructures in YRD and the hydrogen mobility on national expressways. The plan aims to build more than 20 hydrogen expressways and more than 500 hydrogen refuelling stations, with the number of FCVs to reach 20,000 by 2030 in YRD. The total number of refuelling stations in YRD is about 85 at the end of 2023.

The latest regulatory effort made by Shanghai Municipal government is marked by the green hydrogen projected released in ”Shanghai Action Plan to Further Promote New Infrastructure Construction (2023-2026)” in October 2023[61]. The planned hydrogen projects aim to promote the pilot utilisation of hydrogen as energy source in the Lingang (Port Nearby) New District in Shanghai, and to launch a pilot 800MW project for offshore wind power hydrogen production. Two State-owed Enterprises, namely Shenergy Group and Shanghai Electric, bid successfully for the pilot project in 2023. The development trends in Shanghai are shaped by the focus towards off-shore wind RH2 production and the hydrogen-coupled ports.

Among all coastal districts in Shanghai, Lingang New District is emerging as the hydrogen active port area with facilitating policies. In August 2022, the Shanghai NRDC released 22 policies on ”Supporting the High-Quality Development of the Hydrogen Energy Industry in the China (Shanghai) Free Trade Pilot Zone Lingang New District”[62]. In an isolated application scenario such as port, dilemmas such as hydrogen price and infrastructure can be addressed via regulations and incentives. For instance, the port encourages the use of various FCVs by offering fixed rewards for miles. Additionally, subsidies are provided for the market price of hydrogen refuelling, and FCVs are given priority road access rights on designated roads in port area.

#### *4.2.2. Three Provinces: Industrial synergy across hydrogen value chain utilising decentralized by-product hydrogen*

Imagine a paper-fan of YRD, with Shanghai at its pivotal point, the other three provinces including Jiangsu, Zhejiang, and Anhui province unfold like the fan leaves (Fig.17). The national expressways serve as the ribs of the fan. In August 2020, the Central Committee of the Communist Party and the State Council jointly issued the ”Outline for the Integrated Development Plan of the Yangtze River Delta Region”[63], emphasizing the importance of synergy in industries, transportation, and energy infrastructures in YRD.

With 119 (out of 616) chemical industry parks and 3 (out of 7) major chemical bases located in Shanghai, Zhejiang, and Jiangsu, the YRD region accounts for 26% (RMB 2.9 trillion/€377 billion) of the total petrochemical revenue in China in 2020[64]. Considered as a transient approach towards RH2, the industrial by-product hydrogen, with relatively low cost and high availability, is encouraged in the national hydrogen plan as the short-term goal.

In contrast to the landscape of RH2 production in remote natural areas, the by-product hydrogen production is overlapping with the load center area in YRD, which is advantageous in terms of accessibility and decentralized supply. Although the CO2 emission from by-product hydrogen production is not negligible compared with RH2, it is still a significant emission reduction compared with fossil fuel based hydrogen. Estimated in ”China’s Hydrogen Energy Industry Development Report 2020”, subtracting the utilisation by local industries, the by-product hydrogen remains a potential capacity of 4.5 Mt/y, which can supply the operation of 970,000 fuel cell buses annually[65].

[57] People’s Net (Feb 2023), The GDP exceeds RMB 100 billion! 41 cities in the Yangtze River Delta offer outstanding performance in 2022, http://zj.people.com.cn/n2/2023/0208/c228592-40293885.html

[58] People’s Net (Sept 2021), The total permanent population of the Yangtze River Delta reaches 235 million (in Chinese), http://zj.people.com.cn/n2/2021/0930/c186327-34938966.html

[59] Xinhua Net (May 2019), Yangtze River Delta “Hydrogen Corridor” plan is released (in Chinese), http://www.xinhuanet.com/politics/2019-05/24/c_1124539110.htm

[60] Development of Yangtze River Delta Hydrogen Corridor

[61] Shanghai Municipal Government (Oct 2023), Shanghai Action Plan to Further Promote New Infrastructure Construction (2023-2026) (in Chinese), https://www.shanghai.gov.cn/nw12344/20231018/8050cb446990454fb932136c0b20ba4d.html

[62] Shanghai NRDC (Aug 2022), Supporting the High-Quality Development of the Hydrogen Energy Industry in the China (Shanghai) Free Trade Pilot Zone Lingang New District (in Chinese), https://fgw.sh.gov.cn/cmsres/41/41597a91ca63462fa360151f67247c35/af68fe01068d6880f74486e238eddf7f.pdf

[63] NRDC (Aug 2020), Outline for the Integrated Development Plan of the Yangtze River Delta Region (in Chinese), https://www.ndrc.gov.cn/xwdt/ztzl/cjsjyth1/ghzc/202007/t20200728_1234708.html

[64] Souhu News (Jul 2022), Development status and competitive situation of the Yangtze River Delta chemical industry park (in Chinese), https://www.sohu.com/a/565513435_121127986

[65] China EV100 (2020), China’s Hydrogen Energy Industry Development

**Eastern China: Yangtze River Delta Hydrogen Corridor**

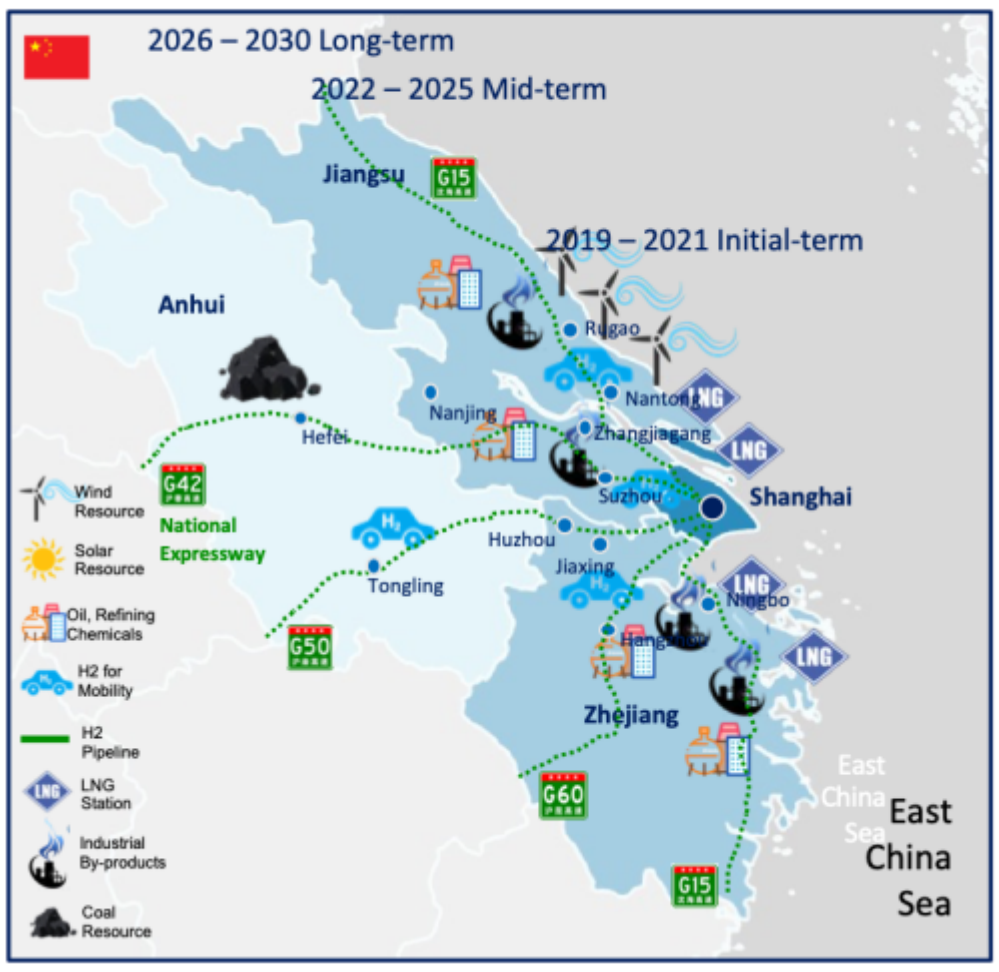

**Development Plan for the Yangtze River Delta Hydrogen Corridor (2020)**

| Developing Phase | No. of H2 refueling stations | No. of fuel cell vehicles | No. of key cities involved | No. of Expressway stations | No. of Expressways covered (H2 Corridor Concept) |
|---|---|---|---|---|---|
| 2019 – 2021 Initial term | > 40 | > 5'000 | 7+ | 8+ | 4 (G15, G42, G50, G60) |
| 2022 – 2025 Mid-term | > 200 | > 50'000 | 20+ | 36+ | 10+ |
| 2026 – 2030 Long-term | > 500 | > 20'000 | 40+ | 48+ | 20+ |

**Current Progress of H2 Refueling Stations in YRD H2 Corridor**

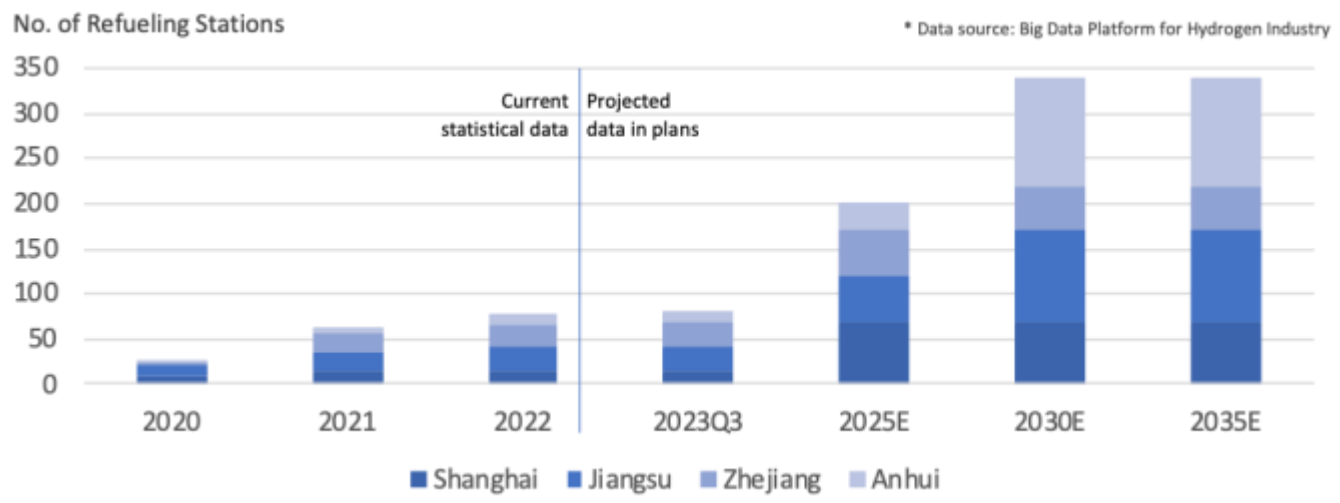

Figure 17: Overview of Yangtze River Delta hydrogen corridor and development plan (2020) [60]

According to Jiangsu Provincial NRDC, the industrial by-product hydrogen (e.g. chlor-alkali exhaust, propane dehydrogenation, ethane cracking) production capacity is about 88,000 ton/y and the grey hydrogen from steel industry (e.g. coke oven gas) is about 188,000 ton/y[66]. Across the hydrogen value chain, Jiangsu Province has also a strong industrial presence in hydrogen storage, infrastructure, and applications including mobility and power electricity. This industrial pattern is advantageous and can accelerate the establishment of hydrogen eco-systems, as envisaged in the ”14th Five-Year Plan for the Development of the New Energy Vehicle Industry” in Jiangsu Province[67].

The cooper-competitive Zhejiang province, with a strong industrial focus on hydrogen related equipment beyond chemical plants and mobility, issued ”Hydrogen Energy Equipment Industry Development Action Plan (2023-2025)” in September 2023 [68]. The local government aims to support Jiaxing City for construction of YRD (Jiaxing) Hydrogen Industrial Park[69], Zhoushan Archipelago for hydrogen equipment island in the sea[70], cultivating advanced manufacturing clusters in the region.

Anhui province, the energy hub in YRD region, owns abundant natural resources including coal, minerals, and nonferrous metals in eastern China. In addition to the natural resource based chemical and steel industries, the nonferrous metals (e.g. Tongling Nonferrous) are key catalysts to the scaling manufacturing of electrolysers and FCVs[71]. Notably, hydro-power is recognized as a promising renewable source for RH2 production in Lu’an City with a potential capacity of 534 MW[72].

In bird’s view of the YRD hydrogen industry landscape, each province and city possesses its gifted natural resources and focused industry clusters. The logistics from Shanghai and port to three provinces, and the energy and materials supply from Anhui to other areas, together are transforming the YRD into an enormous Hydrogen-Industry-Transport Nexus. The challenge lies in how to complement each other’s advantages, enhancing industrial synergies, develop differential road-maps, and accelerate the integration of the YRD industries.

### 4.2.3. *Hydrogen Expressways: Integrating hydrogen industrial clusters via cross-provincial logistics*

The concept of the YRD hydrogen corridor derived from the hydrogen expressways for road transportation that interconnects Shanghai with other three provinces in YRD towards cross-provincial integration. As depicted in Fig.17, more than 20 national expressways are expected to be equipped with hydrogen refuelling stations in YRD by 2030 to enable the ”mileage anxiety” free hydrogen logistics. In initial develop-

Report 2020 (in Chinese), https://www.sdtzsb.com/Uploads/editor/6095e307b4c84.pdf

[66] NRDC Jiangsu Province (Jun 2023), Reply: Optimize and promote the high-quality development of Jiangsu Province’s hydrogen energy industry (in Chinese), https://fzggw.jiangsu.gov.cn/art/2023/7/14/art_64419_11023074.html

[67] Government of Jiangsu Province (Jan 2022), ”14th Five-Year ” Plan for the Development of the New Energy Vehicle Industry (in Chinese), https://www.js.gov.cn/art/2022/1/21/art_64797_10324243.html

[68] Zhejiang Provincial Department of Economy and Information Technology (Sept 2023), Hydrogen Energy Equipment Industry Development Action Plan (2023-2025) (in Chinese), https://www.autoreview.com.cn/show_article-18948.html

[69] Jiaxing Daily (Sept 2023), Jiaxing Port Area strives to build a hydrogen energy industry cluster in the Yangtze River Delta (in Chinese), https://www.kczg.org.cn/rules/detail?id=5802264

[70] Government of Zhoushan (Sept 2020), Zhoushan City Guidelines for Accelerating the Cultivation of Hydrogen Energy Industry Development (Draft for Comments) (in Chinese), http://www.xch3.com/Resource/BPPfile/2020-09-29/20200929162055197.pdf

[71] Government of Tongling (Jun 2023), Reply: Suggestions on accelerating the high-quality development of our city’s hydrogen energy industry (in Chinese), https://www.tl.gov.cn/openness/OpennessContent/show/921320.html

[72] Government of Lu’an (Sept 2020), Lu’an City Hydrogen Energy Industry Development Plan (2020-2025) (in Chinese), https://www.ahjinzhai.gov.cn/group3/M00/0F/82/wKgSG1_iyRWAFJ2vABOz9piaT_4478.pdf

ment phase, it focuses on application scenario that supplying buses, logistic trucks, and taxis etc.

Zhejiang is leading the constructions of hydrogen expressways from Shanghai, to Jiaxing and Ningbo City along the Expressway G15 (Shenyang-Haikou), as proposed in the development plan of YRD hydrogen corridor. In Oct 2022, Sinopec started the operation of the first refuelling super-station that offers gasoline, diesel, e-charging piles, LNG, and hydrogen. It offers hydrogen refilling for both 35 MPa and 70 MPa hydrogen FCVs at the same time[73]. This super-refuelling station enables 50 heavy-duty fuel cell trucks deployed at the harbour area of Jiaxing City in Zhejiang, to transport goods freely on the national expressway. Notably, this project is marked as the only profitable project in YRD as of 2023Q1, with a local hydrogen price of about 20 RMB/kg (€2.5/kg).

In contrast to the pipeline, which delivers hydrogen from centralized production sites continuously to centralized immobilised end-users, the concept of hydrogen expressways offers high adaptability of existing road infrastructures to enabling the decentralised hydrogen production and versatile hydrogen mobility. Limited by the high costs of refuelling station constructions, at present, this concept is mainly serving for heavy-duty logistics.

### *4.3. Northern Coastal Europe - Delta Rhine Corridor for Hydrogen and CO2*

#### *4.3.1. Dutch Coastal Region: Rotterdam off-shore wind hydrogen production and seaport adaptation for hydrogen hub*

The Port of Rotterdam, as the largest seaport in Europe, is a central hub of conventional and renewable energy. Remarkably, the annually transported energy (8,800 PJ) through this port is three times more than the Netherlands' total energy consumption, and 13% of the energy demand in Europe[74]. Seeking to be the world leading port for sustainable energy, the port authorities are in collaboration with multiple private and public partners to establish an extensive hydrogen network within the port complex. The goal is to transform Rotterdam Port into a global center for hydrogen production, import, usage, and distribution to neighbouring countries in Northwest Europe.

In May 2022, the port of Rotterdam, the local government, and industrial agglomerates together issued a statement that they will work jointly to supply Europe with at least 4.6 Mt/y of hydrogen by 2030[76]. Leveraging the abundant wind resources along the coast, the Netherlands issued its national hydrogen strategy in March 2020, aimed for an electrolyser capacity of 500 MW by 2025 and a target of 3-4GW by 2030[77]. Both the port of Rotterdam and the Netherlands' government have rolled out clear goals for scaling up the RH2 production mobilising local industry agglomerates and renewable resources.

As a cornerstone of Rotterdam port, H2-Fifty project was kicked off in Feb 2022 with a jointly agreement signed by HyCC and BP. The project aims to build a 250 MW RH2 plant in Rotterdam's port area, producing up to 20,000 - 30,000 tons H2/y[78]. This already accounts for half of the goals outlined by the government in national hydrogen strategy. With massive RH2 production capacity by 2025, the port will be able to replace all grey hydrogen imported today by the BP Rotterdam Refinery as feedstock, resulting an emission reduction up to 270,000 ton/y. The project has also been granted IPCEI status by the European Commission later in September 2022.

#### *4.3.2. German Rhine Region: Demand center for hard-to-abate sectors including steel and chemicals*

The energy supply from the port of Rotterdam is not just limited in the Netherlands. The German Rhine River Region, consisting two federal states namely North Rhine-Westphalia (NRW) and Rhineland-Palatine (RLP), is receiving goods and energy from the port continuously for feeding its local industries. The federal state of North Rhine-Westphalia is Germany's most populous state with the highest GDP. The heart of German industry, the Ruhr Area, is also located in the center of NRW with numerous heavy, manufacturing, chemical industry agglomerates. Towards the upstream of the Rhine River, the state of Rhineland-Palatine boards with multiple countries and is the home to the largest chemical base operated by BASF and Germany's largest research-based pharmaceutical company, Boehringer Ingelheim.

Germany, with leading number of refuelling stations in the EU, is actively working on hydrogen regulations and national strategy. The government issued as early in Jun 2020 the first national hydrogen strategy, aimed for 5 GW of electrolyser capacity by 2030 and 10 GW by 2040 in Germany, with a strong focus of low-carbon hydrogen[79]. The newly updated national hydrogen strategy in July 2023 outlined the import strategy of hydrogen and its derivatives, as well as the development plan for national hydrogen infrastructures[80].

The two states in Rhine Region issued their own hydrogen roadmap in response to the national hydrogen strategy since 2020. For instance, in NRW, beyond the goals for emission reduction, it also specified goals for hydrogen mobility and infrastructures by 2025. NRW aims to have more than 400 fuel cell trucks, at least 20 truck filling station, 60 car filling stations, 500 fuel cell buses, and the first hydrogen-powered barges. It also outlines the goal of 120 km hydrogen pipelines and more than 100 MW electrolysis plants for industrial hydrogen production by 2025[81]. In RLP, the roadmap envisaged an expansion of solar and wind power by at least 500 MW/y by 2030. Approximately one-eighth of the hydrogen demand will be met through electrolysis capacity in RLP itself, with a capacity of about 1.3 GW[82].

[73]Souhu News (Oct 2022), Sinopec's first wholesale and retail "five-in-one" integrated energy service station officially opened (in Chinese), https://www.sohu.com/a/592287806_121123883

[74]Port of Rotterdam (May 2020), Port of Rotterdam Becomes International Hydrogen Hub, https://www.portofrotterdam.com/sites/default/files/2021-06/hydrogen-vision-port-of-rotterdam-authority-may-2020.pdf

[75]Development of Delta Rhine Corridor for Hydrogen

[76]Port of Rotterdam (May 2022), Rotterdam to Supply Europe with 4.6 Mt of Hydrogen by 2030, https://www.portofrotterdam.com/sites/default/files/2022-05/Rotterdam%204.6%20Mton%20hydrogen%20in%202030%20proposition.pdf

[77]Government of Netherlands (Mar 2020), Government Strategy on Hydrogen, https://www.government.nl/documents/publications/2020/04/06/government-strategy-on-hydrogen

[78]H2-Fifty Rotterdam (Dec 2023), Project Vision of H2-Fifty, https://www.h2-fifty.com/about

[79]BMBF (Jun 2020), The National Hydrogen Strategy, https://www.bmbf.de/bmbf/shareddocs/downloads/files/bmwi_nationale-wasserstoffstrategie_eng_s01.pdf?__blob=publicationFile&v=2

[80]BMWK (Jul 2023), National Hydrogen Strategy Update, https://www.bmwk.de/Redaktion/EN/Publikationen/Energie/national-hydrogen-strategy-update.pdf?__blob=publicationFile&v=2

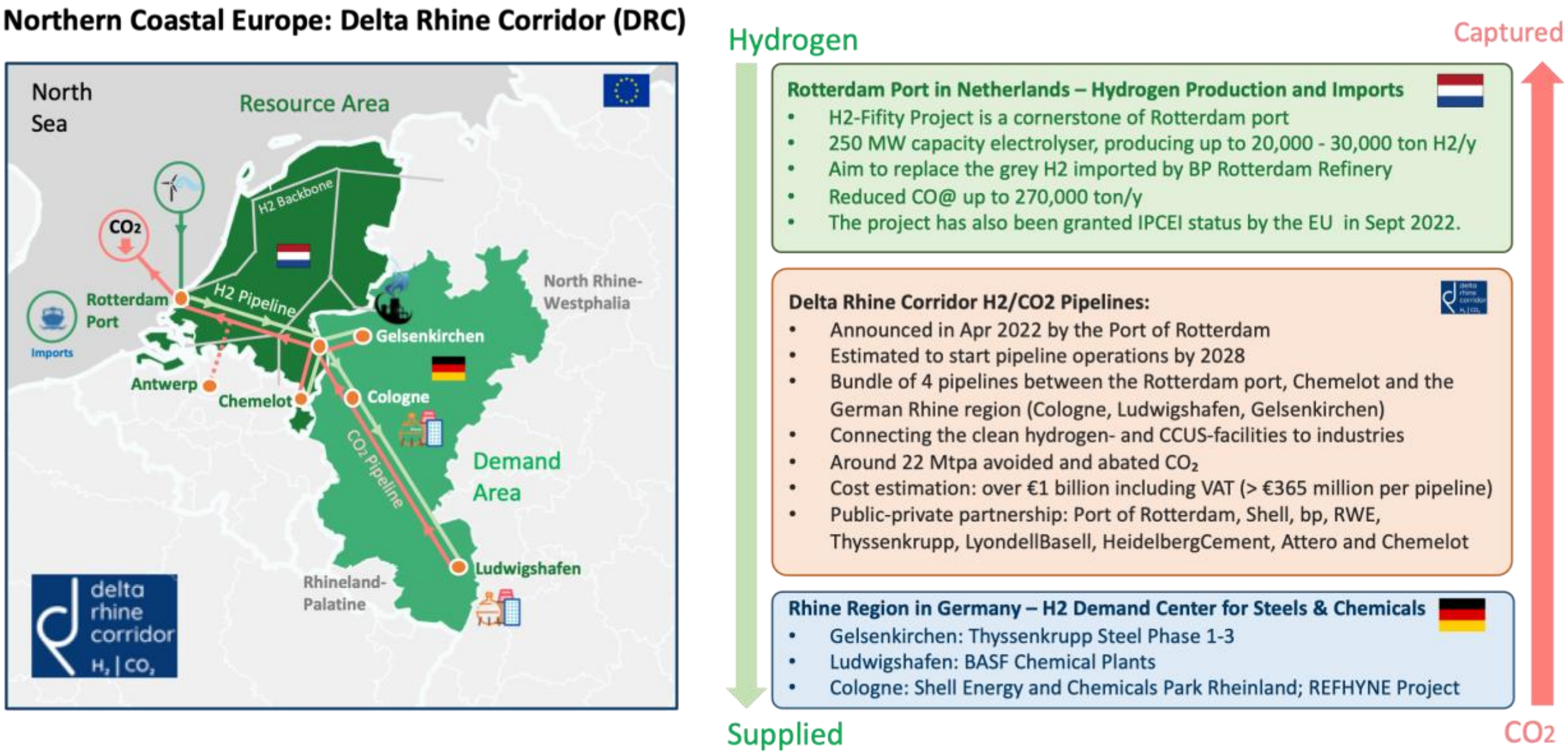

Figure 18: Overview of Delta Rhine Corridor with hydrogen and CO2 pipelines (2022) [75]

The ambitious goals of hydrogen mobility and infrastructure show that, beyond the existing hard-to-abate sectors such as steel and chemicals in the Rhine Region, there will be increasing demand from mobility as well, transforming this area into heavy load center of northwestern Germany.

Currently, Thyssenkrupp Steel in Gelsenkirchen (Duisburg, NRW), is a major consumer of green electricity and hydrogen. Owning large steel production plan in NRW, Thyssenkrupp Steel accounts for nearly 2.5% of the total CO2 emission in Germany[83]. Similarly, BASF operates the largest chemical base in Ludwigshafen in RLP, resulting to 18.4 Mt/y CO2 emission[84], comparable with 20 Mt/y from Thyssenkrupp Steel. In both scenarios, the large consumption of hydrogen and the CO2 emission reduction have to be tackled.

*4.3.3. DRC Hydrogen/CO2 Pipelines: Supplying RH2 from Rotterdam to Rhine region and capturing CO2 for storage in North Sea*

In Europe, Germany and the Netherlands are aiming for carbon neutrality by 2045 and 2050, respectively. The port of Rotterdam is serving the industries not only in Netherlands, but also extensively in northwestern Germany. To tackle the dual-challenges of hydrogen supply and CO2 reduction spontaneously in German and Dutch heavy industries, the Delta Rhine Corridor (DRC) Joint Project (Fig.18) was released in Jun 2021 by the port of Rotterdam[85]. The goal of the DRC project is to recycle CO2 from German and Dutch industries, then capture and storage CO2 in the North Sea. In parallel, industries in DRC will be supplied with low-carbon hydrogen produced in or imported to Rotterdam port[86].

Beyond the support from public authorities in DRC, private companies including BASF, Gasunie, OGE and Shell have jointed as a force to advance the DRC project. It is expected to transport H2 from 2026 and CO2 as of 2028. Currently, the plan is to connect the port of Rotterdam with Chemelot (Dutch chemical industry area), Gelsenkirchen (e.g. Thyssenkrupp Steel), Cologne (e.g. Shell Energy and Chemical Park), and finally reaching Ludwigshafen (e.g. BASF Chemical base) with both hydrogen and $CO_2$ pipelines. Once the DRC project is commissioned, it is expected to avoid and abate around 22 Mt/y $CO_2$, which is roughly 2.3% of the total $CO_2$ emission in Germany in 2022.

[81] Ministry of Economic Affairs, Innovation, Digitalization and Energy of NRW (2020), Hydrogen Roadmap North Rhine-Westphalia, https://www.wirtschaft.nrw/sites/default/files/documents/mwide_br_wasserstoff-roadmap-nrw_eng_web.pdf

[82] Ministry for Climate Protection, Environment, Energy and Mobility Rhineland-Palatinate (Nov 2022), Wasserstoffstudie mit Roadmap Rheinland-Pfalz (in German), https://mkuem.rlp.de/fileadmin/14/Themen/Energie_und_Klimaschutz/6._Gruener_Wasserstoff/Wasserstoffstudie_mit_Roadmap_Rheinland-Pfalz.pdf

[83] Thyssenkrupp Steel (May 2022), H2BF Phase 1: Wasserstoffeinblasversuche am Hochofen 9 der thyssenkrupp Steel Europe AG (in German), https://www.vdkf-ev.de/web/site/wp-content/uploads/2022/05/2-Spatzker_VDKF_H2BF-Phase-1.pdf

[84] BASF (2022), BASF Report 2022: Energy and Climate Protection , https://report.basf.com/2022/en/managements-report/sustainability-along-the-value-chain/safe-and-efficient-production/energy-and-climate-protection.html

[85] Port of Rotterdam (Jun 2021), Feasibility Study for Port of Rotterdam - Chemlot - North Rhine-Westphalia Pipelines, https://www.portofrotterdam.com/sites/default/files/2021-06/202104id-095_delta_corridor_en.pdf

[86] Delta-Rhine-Corridor (Dec 2023), Cross-border pipeline infrastructure - supporting the EU's climate goals, https://www.delta-rhine-corridor.com/en

The counter flow of $H_2/CO_2$ is an unique approach that bridges the supply and demand gap of RH2 in DRC, at the same time, directs the $CO_2$ emission to storage sites in the sea. EU sets the goal for GHG reduction by at least 55% by 2030, thus this novel design serves as a role model that actively captures the $CO_2$ emission from heavy industries and provides flexibility for industrial adaptions in the context of energy transition towards hydrogen economy.

## 5. Summary and Outlook

**An emerging dynamic economy fueled by hydrogen and its applications.** In the micro-realm, the exergy of the hydrogen molecule is high, as is the entropy of the hydrogen economy in the macro-world. Across the Eurasian continent, both China and the EU are targeting hydrogen development and racing towards carbon-neutral goals by 2050 for the EU and 2060 for China. Despite the socio-economic turbulence during the COVID-19 pandemic, both economic powerhouses issued 10-15 year hydrogen strategies between 2020 and 2022, kicking off the race towards a hydrogen-powered future economy.

**China's risk-averse strategy utilizing transitional carriers.** At present, China's national hydrogen strategy prioritizes establishing industrial synergies and competencies across the full hydrogen value chain, from production, transport, and storage to diversified applications. Provinces and regions have rolled out even more ambitious hydrogen production and mobility targets, leveraging local renewable resources and facilitating industry agglomerations. Legislative efforts in the Energy Law (draft) aim to classify hydrogen as an energy carrier rather than a hazardous chemical, with the goal of easing restrictions on its production, transport, and storage. In practice, China adopts transitional carriers, including by-product hydrogen and hybrid mobility, as well as demo-trial projects, indicating a risk-averse strategy. The central government has underscored the importance of "quality" over "quantity" in the development of the hydrogen industry, signaling a need for caution in the policy decisions of local governments.

**A mixed bag of progress in H2 value chain and concerns in sustainability for China.** In the last 10 years, significant progress has been achieved across the H2 value chain in China. Inner Mongolia is leading RH2 production with mixed renewables through the commissioning of MW electrolysis projects. The national "West-to-East" hydrogen pipeline from Inner Mongolia to Beijing is expected to bridge the RH2 gap by supplying the Capital Economic Circle (CEC), which is the load and utility center of Northern China. The scaling deployment of fuel cell vehicles (FCVs) in both the summer and winter Olympic Games in Beijing clearly shows the country's ambition in hydrogen mobility, highlighting the emerging demand for RH2. Navigating to the load center of Eastern Coastal China, the Yangtze River Delta (YRD) possesses advantageous but diversified industries across the hydrogen value chain, with decentralized and accessible low-cost industrial by-product hydrogen supplied via and used for hydrogen expressways. State-owned enterprises (SOEs) are leading the upstream of the hydrogen value chain through rapid deployments of demonstration projects and large-scale production plants, whereas private companies are concentrated towards the downstream and motivated for scaling up various applications. In the short term, economic benefits are not the decisive factors for the initiation of most SOE-led projects, leading to rapid final investment decisions (FID) and commissioning of large projects compared to Europe. However, this approach introduces uncertainties regarding the aftermath and continuity of DEMO projects in the long run, particularly when future subsidies are gradually withdrawn. Utilizing by-product hydrogen is pragmatic during transition periods, but it diverges from China's climate commitment and environmental goals and hinders the market penetration of RH2 unless it achieves competitive LCOH. Last but not least, long-distance hydrogen transmission via electron or molecule remains a challenge in the techno-economic roadmap.

**EU's hybrid strategy targeting sustainable scaling-up and de-risking imports.** In contrast, the current EU-level hydrogen strategy prioritizes scaling up electrolyzer installations, aiming for the production of 10 Mton RH2 by 2030. The EU has also clarified its import goal of an equivalent amount, recognizing that achieving self-sufficient RH2 production is unrealistic and that a hybrid approach is necessary. Thus, it is planning diversified imports from North Africa and the Middle East to de-risk the energy supply chain. The EU has also enacted a few mandatory regulations to achieve this goal: transitioning 42% of industrial hydrogen to RFNBOs by 2030 and mandating that 5.5% of the total energy quota in the transport sector (including vehicles, ships, aircraft, etc.) be biofuels or RFNBOs by 2030. To increase the supply and create a market for hydrogen applications, the EU has established the world's first European Hydrogen Bank (EHB) with a €3 billion budget for unlocking private investments, motivating stakeholders within and beyond Europe. Additionally, the Clean Hydrogen Partnership is a unique public-private partnership supporting R&D in hydrogen technologies as of 2021. The hybrid strategy in combination of domestic production and imports and legislative efforts require good coordination between global entities and timely regulatory support from governments.

**A tapestry of advanced technologies and landing challenges during EU's H2 development.** The EU's early-stage R&D of hydrogen and fuel cell technologies can be traced back to the 80s. Currently, hydrogen represents about 2% of the EU's energy mix, which is expected to account for up to 20% by 2050. Towards a climate-neutral Europe, pioneering EU countries such as Germany and the Netherlands have formed a joint force in the Delta Rhine Corridor (DRC), unleashing the capacity of offshore wind RH2 produced in Rotterdam Port and transported through hydrogen pipelines to the demand center in the Rhine Region, which hosts hard-to-abate German industries including steel and chemicals. Uniquely, there are $CO_2$ counter-flow pipelines constructing in parallel, aiming to recycle $CO_2$ before emission and later captured and stored in the North Sea. Multi-national companies are in collaboration with local governments and agencies in major projects, strengthening the public-private partnerships. However, as reported by the Hydrogen Council in 2023, only 4% of announced investments in clean hydrogen in Europe have passed FID, which is significantly lower than North America (15%) and China (35%). Besides, the hydrogen technology roadmap (e.g., PEM electrolyser, RH2) adopted by the EU is on the novel side compared to China (e.g., ALK electrolyser, LCH2), leading to huge challenges in scaling up with cost reduction.

**Lessons and reflections from the trajectory of EV development.** At the 2024 World Economic Forum in Davos, the German Technology Trade Association (VDE) highlighted the pressing concern that the Chinese domination of the solar and battery industries could be repeated with electrolysers. Reflecting on the past decade, the Chinese government persistently provided over 200 billion RMB (25.8 billion EUR) as subsidies for the purchase of at least three million EVs (2010-2020) along with a tax break policy. In comparison, the German government has offered so far 9.5 billion EUR subsidizing the purchase of two million EVs by the end of 2023. With rapid product iteration cycles (12 to 24 months) and low-cost manufacturing capabilities in China, the European EV market is facing challenges that demand expedited decision-making processes and reductions in legislative costs. The termination of China's EV subsidy program in 2022 marked a turning year, overlapping with the ramp-up in subsidies for the hydrogen industry across the entire value chain in China, though initially not at the same level as the EV subsidies.

**A recommended Tian Ji's horse racing strategy**. Being the second-largest trading partners to each other, China and the EU are essentially at different development stages in the modern era, with notable contrasts in demographics and political systems. The question is, what sits on two sides of the coin of hydrogen economy. Just as Tian Ji strategically deployed his weakest horse to win the race with the King of Qi-Kingdom in ancient China, each region should capitalize on its strengths while acknowledging its limitations compared to the competitors, considering the entire hydrogen value chain. In response to the government policy, SOEs in China can mobilize abundant resources in a short time scheme and initialize large projects beyond prioritizing short-term economic profits. This is not essentially the case in Europe where multinational enterprises are calculating both short and long-term benefits for kicking off projects, alongside waiting for FID and subsidies to be in place. On the other hand, Europe has been leading the R&D of advanced electrolyzer and fuel cell technologies with patent barriers for decades, as well as progressive hydrogen standards and certifications, asserting an advantageous position in high added-value for the supply chain. Deploying top-tier horses across all segments of the value chain is simply impossible, thus both China and Europe should identify their suitable and practical positions and complement each other in the race for a shared future.

## Declaration of competing interest

The authors declare that they have no known competing financial interests or personal relationships that could have appeared to influence the work reported in this paper.

## License